\documentclass[]{raa_twocolumn}

\usepackage{graphicx}
\usepackage{amssymb,amsmath}
\usepackage{txfonts}
\usepackage{natbib}
\usepackage{threeparttable, tablefootnote}
\bibpunct{(}{)}{;}{a}{}{,} 
\usepackage[colorlinks = true, citecolor = blue]{hyperref}
\renewcommand*{\tablefootnote}\textsuperscript{{\alph{footnote}}}

\usepackage{lscape}
\begin{document}

   \title{The composite X-ray spectra of radio-loud and radio-quiet SDSS quasars
}

   \volnopage{Vol.0 (20xx) No.0, 000--000}      
   \setcounter{page}{1}          

   \author{Minhua Zhou
      \inst{1,2}
   \and Minfeng Gu
      \inst{1}
   }

   \institute{Key Laboratory for Research in Galaxies and Cosmology, 
   	Shanghai Astronomical Observatory, Chinese Academy of Sciences, 
   	80 Nandan Road, Shanghai 200030, China; {\it zhoumh@shao.ac.cn}\\
   	\and
   	University of Chinese Academy of Sciences,
   	19A Yuquan Road, Beijing 100049, China; {\it gumf@shao.ac.cn}\\
\vs\no
   {\small Received~~20xx month day; accepted~~20xx~~month day}}

\abstract{We present the study on the X-ray emission for a sample of radio-detected quasars constructed from the cross-matches between SDSS, FIRST catalogs and XMM-Newton archives. A sample of radio-quiet SDSS quasars without FIRST radio detection is also assembled for comparison. 
We construct the optical and X-ray composite spectra normalized at rest frame $4215\,\rm \AA$ (or $2200\,\rm \AA$) for both radio-loud quasars (RLQs) and radio-quiet quasars (RQQs) at $z\le3.2$, with matched X-ray completeness of 19\%, redshift and optical luminosity. 
While the optical composite spectrum of RLQs is similar to that of RQQs, we find that RLQs have higher X-ray composite spectrum than RQQs, consistent with previous studies in the literature. By dividing the radio-detected quasars into radio loudness bins, we find the X-ray composite spectra are generally higher with increasing radio loudness. 
Moreover, a significant correlation is found between the optical-to-X-ray spectral index and radio loudness, and there is a unified multi-correlation between the radio, X-ray luminosities and radio loudness in radio-detected quasars. These results could be possibly explained with the corona-jet model, in which the corona and jet are directly related.
\keywords{methods: statistical --- catalogs --- quasars: general --- X-rays: general}
}

   \authorrunning{M. Zhou \& M. Gu }            
   \titlerunning{Composite X-ray spectrum}  

   \maketitle

%
%
\section{Introduction}           
\label{sect:intro}

It is well known that active galactic nuclei (AGNs) can be classified into radio-loud or radio-quiet AGNs based on radio loudness \cite[defined as the ratio of the radio to optical luminosity, $R=f_{5\rm GHz}/f_{4400\rm \AA}$,][]{1989AJ.....98.1195K}, with $R\ge 10$ and $R<10$ in the former and the latter, respectively. The main difference between two populations is the presence of powerful relativistic jet in radio-loud AGNs (RL-AGNs), while the jet is thought to be weak or even absent in radio-quiet AGNs (RQ-AGNs) \citep{1993ARA&A..31..473A,1995PASP..107..803U}. 

The systematic comparison studies on the spectral energy distributions (SEDs) in RL-AGNs and RQ-AGNs have been carried out \citep[e.g.,][]{1994ApJS...95....1E,2011ApJS..196....2S}. It has been found that the composite SED of RL-AGNs is similar to that of RQ-AGNs in infrared and optical/UV bands, but they have significant differences in radio and X-ray bands. When normalized at optical band, RL-AGNs are more luminous than RQ-AGNs in both radio and X-ray bands \citep[e.g.,][]{1994ApJS...95....1E,2011ApJS..196....2S}. Moreover, RL-AGNs have slightly flatter X-ray spectra \citep{1997MNRAS.288..920L} and weaker reflection components \citep{1998MNRAS.299..449W,2000ApJ...537..654E} compared to RQ-AGNs. While the difference in radio band could be most likely due to the presence of powerful relativistic jets in RL-AGNs, the reason for the difference in the X-ray band is still under debate \citep[e.g.,][]{1987ApJ...313..596W, 1987ApJ...323..243W, 2002MNRAS.332L..45B, 2014ARA&A..52..529Y, 2018MNRAS.480.2861G, 2020MNRAS.492..315G}. 

RL-AGNs may have additional UV and X-ray flux associated with the radio jet \citep[e.g.,][]{1987ApJ...313..596W,1987ApJ...323..243W,2011ApJ...726...20M}. Alternatively, \cite{2002MNRAS.332L..45B} argued that accretion discs in RL-AGNs could be, on average, more ionized than in RQ-AGNs.  
Recently, \cite{2018MNRAS.480.2861G} studied the hard X-ray emission of two carefully selected RL- and RQ-AGN samples with comparable ranges of black hole mass and Eddington ratio. The authors found RL- and RQ-AGNs have similar X-ray spectral slopes and suggested that the hard X-rays in RL-AGNs and RQ-AGNs are likely produced in the same region and by the same mechanism. This result was further supported by the similar distribution of X-ray loudness between Type 1 and Type 2 RL-AGNs \citep{2020MNRAS.492..315G}. 

In our previous work, we revisited the difference of composite X-ray spectrum between radio-loud quasars (RLQs) and radio-quiet quasars (RQQs) by building a new composite X-ray spectrum for a sample of 3CRR quasars \citep{2020ApJsubmitted}. Due to the low-frequency selection, 3CRR sample is dominated by steep-spectrum sources, therefore their SEDs are less likely dominated by beamed jet. By excluding blazars, we found that the X-ray composite spectrum of 3CRR quasars is similar to that of RLQs in \citet[hereafter S11]{2011ApJS..196....2S}, then significantly differs from that of RQQs. 
We proposed that the jet emission at X-ray band in RLQs may be related to the difference of composite X-ray
spectrum between RLQs and RQQs. 

In view of the inconclusive reason for the X-ray difference in RL- and RQ-AGNs, in this work, we further study the X-ray emission of RLQs by constructing the optical and X-ray composite spectra normalized at optical band for a large sample of RLQs and RQQs utilizing the most updated XMM-Newton X-ray catalog. 
In general, RLQs tend to have redder optical spectra than RQQs \citep[e.g.,][]{2002AJ....124.2364I, 2016ApJ...818L...1S}, which could be either due to stronger dust extinction, or intrinsically redder spectra in radio selected quasars. The bias introduced by this effect needs to be considered when selecting and comparing RLQs and RQQs \citep[e.g.,][and references therein]{2019SCPMA..6269511C}. 
Our sample is presented in Section \ref{sec:sample}, and the optical and X-ray data are shown in Section \ref{sec:data}. 
In Section \ref{sec:comp}, we construct the composite optical and X-ray spectra for our sample. 
The results are discussed in Section \ref{sec:discc} and summarized in Section \ref{sec:sum}. 

In this work, we adopt the cosmology parameters $H_{0}=70\rm ~km~ s^{-1}~Mpc^{-1}$, $\Omega_{\rm m}=0.3$ and $\Omega_{\Lambda}=0.7$. 
Photon index $\Gamma$ of power-law is defined by 
$A(E) = K E^{-\Gamma}$, where $K$ is photons at $1~\rm keV$ and $E$ is photon energy.
The spectral index $\alpha$ is defined as $f_{\rm \nu} \varpropto \nu^{-\alpha}$ with $f_{\rm \nu}$ being the flux density at frequency $\rm \nu$. 


\section{Sample selection} \label{sec:sample}

To construct the large sample of RLQs and RQQs, we started from the optically selected quasars in the Sloan Digital Sky Survey (SDSS) Data Release 14 Quasar catalog \citep[DR14Q,][]{2018A&A...613A..51P}, which includes spectroscopically confirmed quasars in SDSS-I, II, III and SDSS-IV/eBOSS. 
Firstly, we selected quasars in DR14Q with radio detection by cross matching within $2\arcsec$ radius with Faint Images of the Radio Sky at Twenty-Centimeters (FIRST) Survey Catalog \citep{2015ApJ...801...26H}. 
The FIRST-detected quasars were then cross matched with 3XMM spectral-fit database \citep[XMMFITCAT \footnote{\url{http://xraygroup.astro.noa.gr/Webpage-prodex/index.html}},][]{2015A&A...576A..61C} of 3XMM-DR7 catalog
using a standard $5\arcsec$ matching radius. We prefer to use the data from XMM-Newton because it has large collecting area, and the advantage of usually long and uninterrupted exposures and thus highly sensitive observations. 
Finally, the blazars in the BZCAT catalog \citep{2009A&A...495..691M,2015Ap&SS.357...75M} Edition 5.0.0 were checked, and then excluded to minimizing the significant jet contribution at X-ray band due to strong beaming effect. These steps result in a sample of 361 radio-detected quasars, and we, hereafter, call it sample A. 

To estimate the radio loudness $R$, we converted FIRST 1.4 GHz flux to rest-frame $5~\rm GHz$ flux by assuming the radio spectral index $\alpha=0.5$. With the measured flux at rest-frame $\rm 4400~\AA$ (see Section \ref{subsec:optical}), the quasars in sample A can be classified into 299 RLQs ($R\ge10$) and 62 RQQs ($R<10$). These 62 RQQs are regarded as radio-detected RQQs (RD-RQQs).

To enlarge the RQQs sample, we included the quasars in DR14Q with XMM-Newton detection within $5\arcsec$ radius, and within the FIRST survey area, however without detection in FIRST catalog within $2\arcsec$ radius. 
The upper limit of radio loudness was then calculated with the flux limit of FIRST ($1~\rm mJy$) for these matched quasars. We chose the quasars with the upper limit of radio loudness less than 10, resulting in a sample of 1411 RQQs. The sample of these non-radio detect RQQs (NRD-RQQs) is referred to as sample B. 

\subsection{RLQs and RQQs}

Following S11, the X-ray composite spectrum will be normalized at $4215~\rm \AA$ (or via $2200~\rm \AA$, see Section 4), thus we focus on the optical and X-ray spectral analysis for the sources at $z\le3.2$ to ensure the coverage of rest frame $4215~\rm \AA$ or $2200~\rm \AA$ in SDSS spectra.
The further selection of sources with sufficient X-ray photons ($\ge$ 200 photon counts) is applied to enable the detailed spectral analysis at high significance (e.g., to search soft X-ray excess), which results in 160 RLQs and 40 RD-RQQs in sample A, and 707 NRD-RQQs in sample B.

Since not all quasars in the samples within XMM-Newton footprint are detected in X-rays, the X-ray completeness (i.e., the fraction of X-ray detected sources with photons $\ge$ 200 within 3XMM footprint) needs to be carefully investigated in order to avoid any biases when comparing different samples. For example, if a subsample has higher X-ray completeness, it will contain more X-ray relatively faint sources and, therefor, likely has relatively lower composite X-ray spectrum. The X-ray completeness of RLQs and RQQs are shown in Table \ref{tab:X-comp}. We find that the X-ray completeness of combined RQQs sample (RD-RQQs and NRD-RQQs, $\sim$19\%) is lower than that of RLQs sample ($\sim$24\%). In order to apply the same X-ray completeness, we pick up the top 19\% of RLQs with most X-ray photon counts ($\ge$315). 

\begin{figure}[htb!]
	\centering
	\includegraphics[width=1.0\columnwidth]{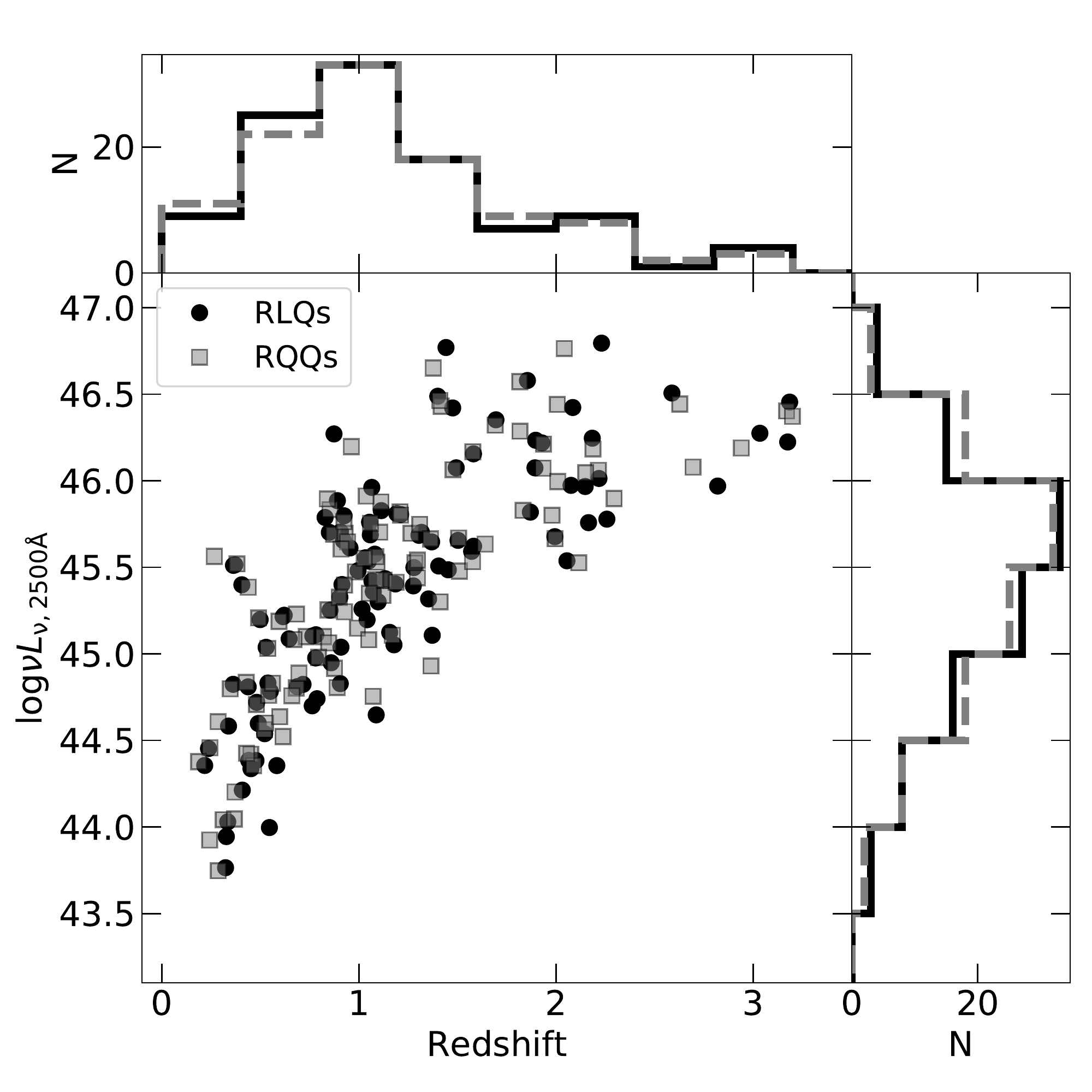}
	\caption{The distribution of redshift and optical luminosity for the matched RLQs and RQQs samples. \label{fig:rl-rq}}
\end{figure}

We noticed that even with the same X-ray completeness, the bias can be still introduced by other selection effects, in which the redshift and optical luminosity are two main issues. 
The redshift and optical luminosity of RQQs are statistically lower than those of RLQs. While the higher redshift in RLQs may cause higher X-ray luminosity thus affecting the composite spectrum, the different optical luminosity distribution may also affect the optical to X-ray slope \cite[see e.g.,][]{2011ApJ...726...20M}. 
To obtain RLQs and RQQs samples with matched redshift and optical luminosity, we performed one-to-one cross-match between RLQs and RQQs by selecting the best matched quasars in $z-\log \nu L_{\nu,\rm2500\AA}$ space within 0.2 unit radius. This results in 106 RLQs and 106 RQQs with matched $z$ and $\log \nu L_{\nu,\rm2500\AA}$ (see Fig. \ref{fig:rl-rq}). 
Among 106 matched RQQs, there are 3 RD-RQQs and 103 NRD-RQQs.
The distributions of X-ray photon in these selected RLQs or RQQs samples are compared with the samples before $z-\log \nu L_{\nu,\rm2500\AA}$ matching. With the Kolmogorov-Smirnov test (KS-test), our selected RLQs and RQQs samples have the same X-ray photon distributions with pre-matched samples with $P-$values of 0.99 and 0.15, respectively. Thus, the same X-ray completeness of our selected RLQs and RQQs may not be affected by the redshift and luminosity matches. The 103 matched NRD-RQQs and all 200 quasars in sample A are shown in Table \ref{tab:sample}.

\begin{table}
	\begin{center}
		\caption[]{The X-ray completeness}\label{tab:X-comp}
		\begin{threeparttable}
		\begin{tabular}{l|ccc}
			\hline\noalign{\smallskip}
			\multicolumn{1}{c}{Sample} & \multicolumn{1}{c}{XMM-FOV} & \multicolumn{1}{c}{N200} & \multicolumn{1}{c}{X-ray completeness} \\
			\multicolumn{1}{c}{(1)} & \multicolumn{1}{c}{(2)} & \multicolumn{1}{c}{(3)} & \multicolumn{1}{c}{(4)} \\
			\hline\noalign{\smallskip}
			NRD-RQQs & 3826 & 707 & 0.18  \\
			RD-RQQs  & 104   & 40  & 0.38  \\
			RQQs     & 3930 & 747 & 0.19  \\
			\hline
			$1\le \log R < 2$ & 328 & 79  & 0.24  \\
			$2\le \log R < 3$ & 257 & 54  & 0.21  \\
			$\log R \ge 3$ & 79 & 27 & 0.34 \\
			RLQs     & 664  & 160 & 0.24  \\
			\noalign{\smallskip}\hline
		\end{tabular}
		\begin{tablenotes}
			\footnotesize
			\item In this Table, Column (1): Samples; Column (2): The count of quasars in 3XMM field of view; Column (3): The quasars with X-ray photons greater than 200; Column (4): The X-ray completeness.
		\end{tablenotes}
	\end{threeparttable}
	\end{center}
\end{table}

\begin{landscape}
	\begin{table}
		\begin{center}
			\caption[]{The Sample}\label{tab:sample}
			\small
			\begin{threeparttable}
				\begin{tabular}{lrrrrrrrrrrccc}
					\hline\noalign{\smallskip}
					\multicolumn{1}{c}{SDSS Name} & \multicolumn{1}{c}{R.A.} & \multicolumn{1}{c}{Dec.} & \multicolumn{1}{c}{$ z $} & \multicolumn{1}{c}{$f_{\rm 1.4 \, GHz}$} & \multicolumn{1}{c}{$R$} & \multicolumn{1}{c}{Plate} & \multicolumn{1}{c}{MJD} & \multicolumn{1}{c}{Fiber} & \multicolumn{1}{c}{XMM ID} & \multicolumn{1}{c}{Counts} & \multicolumn{1}{c}{$\log M_{\rm bh}$} & \multicolumn{1}{c}{$\log L_{\rm bol}/L_{\rm Edd}$} & \multicolumn{1}{c}{Comp.} \\
					\multicolumn{1}{c}{ } & \multicolumn{1}{c}{$deg$} & \multicolumn{1}{c}{$deg$} & \multicolumn{1}{c}{ } & \multicolumn{1}{c}{$\rm mJy$} & \multicolumn{1}{c}{ } & \multicolumn{1}{c}{ } & \multicolumn{1}{c}{ } & \multicolumn{1}{c}{ } & \multicolumn{1}{c}{ } & \multicolumn{1}{c}{ } & \multicolumn{1}{c}{$\rm M_{\odot}$} & \multicolumn{1}{c}{ } & \multicolumn{1}{c}{ } \\
					\multicolumn{1}{c}{(1)} & \multicolumn{1}{c}{(2)} & \multicolumn{1}{c}{(3)} & \multicolumn{1}{c}{(4)} & \multicolumn{1}{c}{(5)} & \multicolumn{1}{c}{(6)} & \multicolumn{1}{c}{(7)} & \multicolumn{1}{c}{(8)} & \multicolumn{1}{c}{(9)} & \multicolumn{1}{c}{(10)} & \multicolumn{1}{c}{(11)} & \multicolumn{1}{c}{(12)} & \multicolumn{1}{c}{(13)} & \multicolumn{1}{c}{(14)} \\
					\hline\noalign{\smallskip}
					\multicolumn{14}{c}{Sample A} \\
					145108.76+270926.9 & 222.787       & 27.157        & 0.065         & 3.43          & 0.59          & 2142          & 54208         & 0637          & 0152660101    & 77010.8       & 8.5           & 0.04          & 0 \\
					135516.56+561244.6 & 208.819       & 56.212        & 0.122         & 6.14          & 6.66          & 8203          & 57428         & 0809          & 0741390401    & 33282.2       & 7.9           & 0.11          & 0 \\
					141700.82+445606.3 & 214.253       & 44.935        & 0.114         & 1.09          & 0.85          & 1287          & 52728         & 0296          & 0109080501    & 30971.1       & 8.0           & 0.11          & 0 \\
					111830.28+402554.0 & 169.626       & 40.432        & 0.155         & 1.04          & 0.41          & 1440          & 53084         & 0204          & 0111290301    & 29577.4       & 8.4           & 0.12          & 0 \\
					151443.07+365050.4 & 228.679       & 36.847        & 0.371         & 48.55         & 38.01         & 1353          & 53083         & 0580          & 0111291001    & 27795.9       & 9.2           & 0.07          & 1 \\
					\multicolumn{1}{c}{...} & \multicolumn{1}{c}{...} & \multicolumn{1}{c}{...} & \multicolumn{1}{c}{...} & \multicolumn{1}{c}{...} & \multicolumn{1}{c}{...} & \multicolumn{1}{c}{...} & \multicolumn{1}{c}{...} & \multicolumn{1}{c}{...} & \multicolumn{1}{c}{...} & \multicolumn{1}{c}{...} & \multicolumn{1}{c}{...} & \multicolumn{1}{c}{...} & \multicolumn{1}{c}{...} \\
					\hline
					\multicolumn{14}{c}{RQQs} \\
					144645.94+403505.7 & 221.691       & 40.585        & 0.267         & 0             & $\le 0.37$          & 1397          & 53119         & 0190          & 0109080601    & 15497.2       & 8.5           & 0.36          & 1 \\
					085841.44+140944.8 & 134.673       & 14.162        & 1.050         & 0             & $ \le 5.12 $          & 5295          & 55978         & 0338          & 0203450101    & 7028.7        & 9.2           & 0.03          & 1 \\
					122018.43+064119.6 & 185.077       & 6.689         & 0.286         & 0             & $ \le 2.07 $          & 1626          & 53472         & 0292          & 0105070101    & 4139.0        & 9.0           & 0.02          & 1 \\
					123335.07+475800.4 & 188.396       & 47.967        & 0.382         & 0             & $ \le 0.97 $          & 6670          & 56389         & 0322          & 0205180301    & 3657.6        & 9.1           & 0.07          & 1 \\
					142417.32+225026.3 & 216.072       & 22.841        & 1.072         & 0             & $ \le 9.03 $          & 6014          & 56072         & 0446          & 0744240101    & 3030.0        & 8.5           & 0.10          & 1 \\
					\multicolumn{1}{c}{...} & \multicolumn{1}{c}{...} & \multicolumn{1}{c}{...} & \multicolumn{1}{c}{...} & \multicolumn{1}{c}{...} & \multicolumn{1}{c}{...} & \multicolumn{1}{c}{...} & \multicolumn{1}{c}{...} & \multicolumn{1}{c}{...} & \multicolumn{1}{c}{...} & \multicolumn{1}{c}{...} & \multicolumn{1}{c}{...} & \multicolumn{1}{c}{...} & \multicolumn{1}{c}{...} \\
					\noalign{\smallskip}\hline
				\end{tabular}
				\begin{tablenotes}
					\footnotesize
					\item In this Table, Column (1): SDSS name (SDSS J); Column (2): Right ascension in degree (J2000); Column (3): Declination in degree (J2000); Column (4): Redshift; Column (5): FIRST flux density in $\rm mJy$; Column (6): Radio loudness; Column ($7-9$): Plate, MJD, and Fiber for SDSS spectra; Column(10): The XMM-Newton observation ID; 
					Column (11): Spectral background subtracted counts from XMMFITCAT; Column (12): Black hole masses; Column (13): Eddington ratio; Column (14): The quasars be used to create the RLQs or RQQs composite X-ray spectrum, 1 for yes, 0 for no.
					Table \ref{tab:sample} is published in its entirety in the machine readable format.
				\end{tablenotes}
			\end{threeparttable}
		\end{center}
	\end{table}
\end{landscape}

\section{Data reduction} \label{sec:data}
\subsection{Optical} \label{subsec:optical}

The SDSS spectra of quasars were analysed with Python-based program PyQSOFit \citep{2018ascl.soft09008G}. The SDSS optical spectra were firstly corrected for Galactic extinction with the reddening map of \cite{1998ApJ...500..525S}, and then shifted to source rest frame by using the redshift in the header of SDSS fits file.
A single power-law and {Fe\,{\footnotesize II}} emission were applied to fit the continuum at several line-free windows \cite[see e.g.,][]{2011ApJS..194...45S}. The flux at $\rm 4400\,\AA$ at source rest frame was then estimated from the fitted power-law continuum, from which the radio loudness can be calculated in combination with the flux at 5 GHz. 

We fitted the continuum-subtracted emission lines with Gaussian profiles \cite[see details in e.g., ][]{2011ApJS..194...45S, 2020MNRAS.491...92L}. All narrow-line components were modeled with a single Gaussian component. The broad components of {H\,{\footnotesize $ \alpha $}}, {H\,{\footnotesize $\beta$}}, {Mg\,{\footnotesize II}}, and {C\,{\footnotesize IV}} were fitted with multiple Gaussian profiles (up to three). 
And then all virial black hole masses were estimated with the empirical relationship between the broad line region radius and the optical/UV continuum luminosity, in combination with the line width of broad emission lines \citep[e.g.,][]{2011ApJS..194...45S}. With different source redshift and availability of emission lines, various lines were used for black hole mass estimation, with {H\,{\footnotesize $ \alpha $}} or {H\,{\footnotesize $\beta$}} for low redshift sources, while {Mg\,{\footnotesize II}} or {C\,{\footnotesize IV}} for high redshift. The bolometric luminosity for our sources were estimated from the corresponding continuum luminosity to emission lines, i.e., $\nu L_{\rm \nu, 5100\AA}$ to {H\,{\footnotesize $ \alpha $}} or {H\,{\footnotesize $\beta$}}, $\nu L_{\rm \nu, 3000\AA}$ to {Mg\,{\footnotesize II}}, and $\nu L_{\rm \nu, 1350\AA}$ to {C\,{\footnotesize IV}} \citep[see e.g.,][]{2011ApJS..194...45S}. The Eddington luminosity was calculated with black hole mass as $L_{\rm Edd}=1.38\times10^{38}(M_{\rm BH}/{\rm M_{\odot}})\,\rm erg\,s^{-1}$. The black hole masses and Eddington ratio are shown in Table \ref{tab:sample}.  

\subsection{X-ray} \label{subsec:x-ray}

For all our sample sources, the XMMFITCAT provides the results of spectral fitting with various simple models (absorbed power-law model, absorbed thermal model, and absorbed black-body model) and complex models (absorbed thermal plus power-law model, absorbed double power-law model, and absorbed black-body plus power-law model). However, we decided to reduce the XMM-Newton data for all quasars by our own rather directly taking from XMMFITCAT, for the sake of spectral analysis in a uniform way. 

We downloaded X-ray data from XMM-Newton Science Archive\footnote{\url{https://www.cosmos.esa.int/web/xmm-newton/xsa}} and processed these data with XMM-Newton Scientific Analysis Software (SAS) package step by step with SAS cookbook\footnote{\url{https://heasarc.gsfc.nasa.gov/docs/xmm/abc/}}. The pn and MOS datasets were processed with $epproc$ and $emproc$ in SAS-15.0.0, and were filtered in energy ranges of $0.2-15.0~\rm keV$ and $0.2-12.0~\rm keV$, respectively. We created and checked the light curve to filter out the time interval with large flares. 
The source X-ray spectrum was then extracted from the pn data from a source-centered radius of $32\arcsec$ with background  source-free region of $40\arcsec$ around the object. If the source locates at the edge of the pn CCD chip or is affected by other source, we accordingly decreased the radius of source region (e.g., $20\arcsec$, or even $15\arcsec$). In the case that the source exactly locates at the gap or is out of pn chips, we extracted the X-ray spectrum from MOS data. 
Pile-up effect was checked with $epatplot$. Finally, all spectra were rebinned with a minimum of 15 counts for background-subtracted spectral channel and oversampling the intrinsic energy resolution by a factor of no larger than three. 

All X-ray spectra were fitted by XSPEC v12.9 with $chi-squared$ statistical method. 
We fitted all spectra with two models, the intrinsic absorbed power-law model with fixed Galactic absorption ($phabs*zphabs*powerlaw$) and the absorbed black-body plus power-law model ($phabs*zphabs*(bbody+powerlaw)$). The black-body component in the second model is used to fit the soft X-ray excess if present.
To search the soft X-ray excess, we firstly fitted the X-ray spectrum in $1.0-10.0~\rm keV$ with a Galactic absorbed \citep{2005A&A...440..775K} single power-law model ($phabs*powerlaw$), and then the residual between the model and data was visually inspected at soft X-ray band $0.3-1.0~\rm keV$. If the residual shows prominent excess, we re-fitted the spectrum with the absorbed black-body plus power-law model. In the cases of weak excess, we fitted the spectrum with both models and the presence of soft X-ray excess will depend on which model better fits the data. When there is no any hints of excess, the intrinsic absorbed power-law model will be used to fit the spectrum.

The results of spectral fit for the final sample sources are given in Table \ref{tab:fitresult}. 

\begin{landscape}
\begin{table}
	\begin{center}
		\caption[]{The results of X-ray spectral fit}\label{tab:fitresult}
		\small
		\begin{threeparttable}
		\begin{tabular}{lcrcccrrrrrrr}
			\hline\noalign{\smallskip}
			\multicolumn{1}{c}{SDSS name} & $\log L_{\rm \nu,2500\AA}$ & \multicolumn{1}{c}{XMM ID} & $\log L_{2-10\rm keV}$ & $\log L_{\rm \nu,2keV}$ & \multicolumn{1}{c}{nH} & \multicolumn{1}{c}{z.nH} & \multicolumn{1}{c}{kT} & \multicolumn{1}{c}{Bbo.norm} & \multicolumn{1}{c}{$\Gamma$} & \multicolumn{1}{c}{Pow.norm} & \multicolumn{1}{c}{$\chi^2_{Red}$} & \multicolumn{1}{c}{d.o.f.} \\
			\multicolumn{1}{c}{ } & \multicolumn{1}{c}{$\rm erg\,s^{-1}\,Hz^{-1}$} & \multicolumn{1}{c}{ } & \multicolumn{1}{c}{$\rm erg\,s^{-1}$} & \multicolumn{1}{c}{$\rm erg\,s^{-1}\,Hz^{-1}$} &  \multicolumn{1}{c}{$10^{20}\rm cm^{-2}$} & \multicolumn{1}{c}{$10^{22}\rm cm^{-2}$} & \multicolumn{1}{c}{$\rm keV$} & \multicolumn{1}{c}{ } & \multicolumn{1}{c}{ } & \multicolumn{1}{c}{ } & \multicolumn{1}{c}{ } & \multicolumn{1}{c}{ } \\
			\multicolumn{1}{c}{(1)} & \multicolumn{1}{c}{(2)} & \multicolumn{1}{c}{(3)} & \multicolumn{1}{c}{(4)} & \multicolumn{1}{c}{(5)} & \multicolumn{1}{c}{(6)} & \multicolumn{1}{c}{(7)} & \multicolumn{1}{c}{(8)} & \multicolumn{1}{c}{(9)} & \multicolumn{1}{c}{(10)} & \multicolumn{1}{c}{(11)} & \multicolumn{1}{c}{(12)} & \multicolumn{1}{c}{(13)} \\
			\hline\noalign{\smallskip}
			\multicolumn{13}{c}{Sample A} \\
			155102.79+084401.1 & 32.29         & 0763780601    & 45.67         & 27.62         & 3.15          & $ 3.13^{+2.95}_{-0.82} $      &           &       & $ 1.56^{+0.29}_{-0.24} $          & 3.04E-05      & 0.83          & 17 \\
			075112.30+291938.3 & 31.63         & 0761510101    & 45.18         & 27.32         & 3.61          & $ \le 0.13 $      & $ 0.11^{+0.03}_{-0.04} $          & 2.36E-06      & $ 2.02^{+0.18}_{-0.09} $          & 1.37E-04      & 1.15          & 38 \\
			111830.28+402554.0 & 29.88         & 0111290301    & 43.90         & 26.14         & 1.45          & $ \le 0.01 $      & $ 0.10^{+0.01}_{-0.01} $          & 1.82E-05      & $ 2.39^{+0.08}_{-0.08} $          & 7.90E-04      & 0.97          & 81 \\
			142129.75+474724.5 & 29.13         & 0094740201    & 44.19         & 26.21         & 1.64          & $ 0.06^{+0.03}_{-0.03} $      & $ 0.07^{+0.01}_{-0.01} $          & 1.45E-04      & $ 1.75^{+0.06}_{-0.06} $          & 3.24E-03      & 1.58          & 106 \\
			145108.76+270926.9 & 29.32         & 0152660101    & 43.45         & 25.72         & 2.78          & $ \le 0.01 $      & $ 0.12^{+0.01}_{-0.01} $          & 5.30E-05      & $ 2.50^{+0.06}_{-0.06} $          & 2.17E-03      & 1.35          & 103 \\
			\multicolumn{1}{c}{...} & \multicolumn{1}{c}{...} & \multicolumn{1}{c}{...} & \multicolumn{1}{c}{...} & \multicolumn{1}{c}{...} & \multicolumn{1}{c}{...} & \multicolumn{1}{c}{...} & \multicolumn{1}{c}{...} & \multicolumn{1}{c}{...} & \multicolumn{1}{c}{...} & \multicolumn{1}{c}{...} & \multicolumn{1}{c}{...} & \multicolumn{1}{c}{...} \\
			\hline
			\multicolumn{13}{c}{RQQs} \\
			002230.48+012218.5 & 30.59         & 0553230201    & 44.92         & 27.10         & 2.82          & $ 0.12^{+0.22}_{-0.12} $      &            &        & $ 2.23^{+0.18}_{-0.16} $          & 1.17E-05      & 1.22          & 43 \\
			004338.10+004615.9 & 30.45         & 0090070201    & 44.74         & 26.76         & 1.79          & $ 0.31^{+1.06}_{-0.31} $      & $ 0.02^{+0.02}_{-0.02} $          & 4.77E-04      & $ 1.75^{+0.42}_{-0.29} $          & 1.11E-05      & 1.10          & 12 \\
			012457.39+015443.0 & 31.13         & 0109860101    & 44.79         & 27.01         & 3.00          & $0.30^{+0.61}_{-0.30}$      &            &        & $ 2.33^{+0.58}_{-0.41} $          & 9.92E-06      & 1.33          & 23 \\
			012507.52-011213.2 & 30.32         & 0743700201    & 44.23         & 26.36         & 4.18          & $\le 0.43$      & $ 0.06^{+0.06}_{-0.03} $          & 4.91E-07      & $ 2.06^{+0.41}_{-0.23} $          & 1.56E-05      & 1.36          & 54 \\
			014022.22-005034.1 & 30.45         & 0744450301    & 44.85         & 27.86         & 2.73          & $ 9.36^{+4.60}_{-6.58} $      & $ 0.05^{+0.02}_{-0.02} $          & 1.02E-04      & $ 2.18^{+1.29}_{-0.62} $          & 4.68E-06      & 0.95          & 14 \\
			\multicolumn{1}{c}{...} & \multicolumn{1}{c}{...} & \multicolumn{1}{c}{...} & \multicolumn{1}{c}{...} & \multicolumn{1}{c}{...} & \multicolumn{1}{c}{...} & \multicolumn{1}{c}{...} & \multicolumn{1}{c}{...} & \multicolumn{1}{c}{...} & \multicolumn{1}{c}{...} & \multicolumn{1}{c}{...} & \multicolumn{1}{c}{...} & \multicolumn{1}{c}{...} \\
			\noalign{\smallskip}\hline
		\end{tabular}
	\begin{tablenotes}
		\footnotesize
		\item In this Table, Column (1): SDSS name (SDSS J); Column (2): The luminosity at rest frame $2500\rm \, \AA$; Column (3): The XMM-Newton observation ID; 
		Column (4): The X-ray luminosity at rest frame $2-10\rm \, keV$; Column (5): The X-ray luminosity at rest frame $2\rm \, keV$; Column (6): Galactic hydrogen column density, in units of $10^{20}~\rm cm^{-2}$; 
		Column (7): The intrinsic hydrogen column density; Column (8$ - $9): parameters of black-body components; Column (10$ - $11): parameters of power-law components; Column(12$ - $13): Reduced $\chi^2$ and degrees of freedom. 
		Table \ref{tab:fitresult} is published in its entirety in the machine readable format.
	\end{tablenotes}
	\end{threeparttable}
	\end{center}
\end{table}
\end{landscape}

\section{Composite spectra} 
\label{sec:comp}

To construct the X-ray composite spectrum, we firstly created the extinction/absorption-corrected optical and X-ray spectra for individual quasars, as S11 did. The optical and X-ray spectra were then normalized at rest-frame $4215~\rm \AA$ for each object. When the rest-frame SDSS spectra cover $4215~\rm \AA$ at low-redshift quasars, the normalization factor
is the mean flux density within $30~\rm \AA$ around $4215~\rm \AA$. While the SDSS spectra do not cover $4215~\rm \AA$ for high-redshift sources, we normalized at $2200~\rm \AA$ to the optical composite spectrum constructed with all spectra 
normalized earlier at $4215~\rm \AA$ in the same sample. In these cases, the normalization 
factor is the mean flux density within $50~\rm \AA$ around 
$2200~\rm \AA$.

To build the composite spectrum for the sample, we binned the normalized optical and X-ray spectra in the same way as \cite{2020ApJsubmitted}, but with different bin size. 
We rebinned the optical spectra in 125 bins from 14.7 to 15.2 in $\log\nu$ space with the bin size $\log\nu=0.004$ Hz. 
Then the median value in each bin was adopted as the flux of the optical composite spectrum. 
We used 14 bins for X-ray spectra from 17.08 to 18.76 in $\log\nu$ space with the bin 
size $\log\nu=0.12$ Hz. 
The X-ray median composite spectrum was then obtained from the median $\log \nu{f}\sb\nu$ values in each bin. 

\begin{figure*}[htb!]
	\centering
	\includegraphics[width=1.0\textwidth]{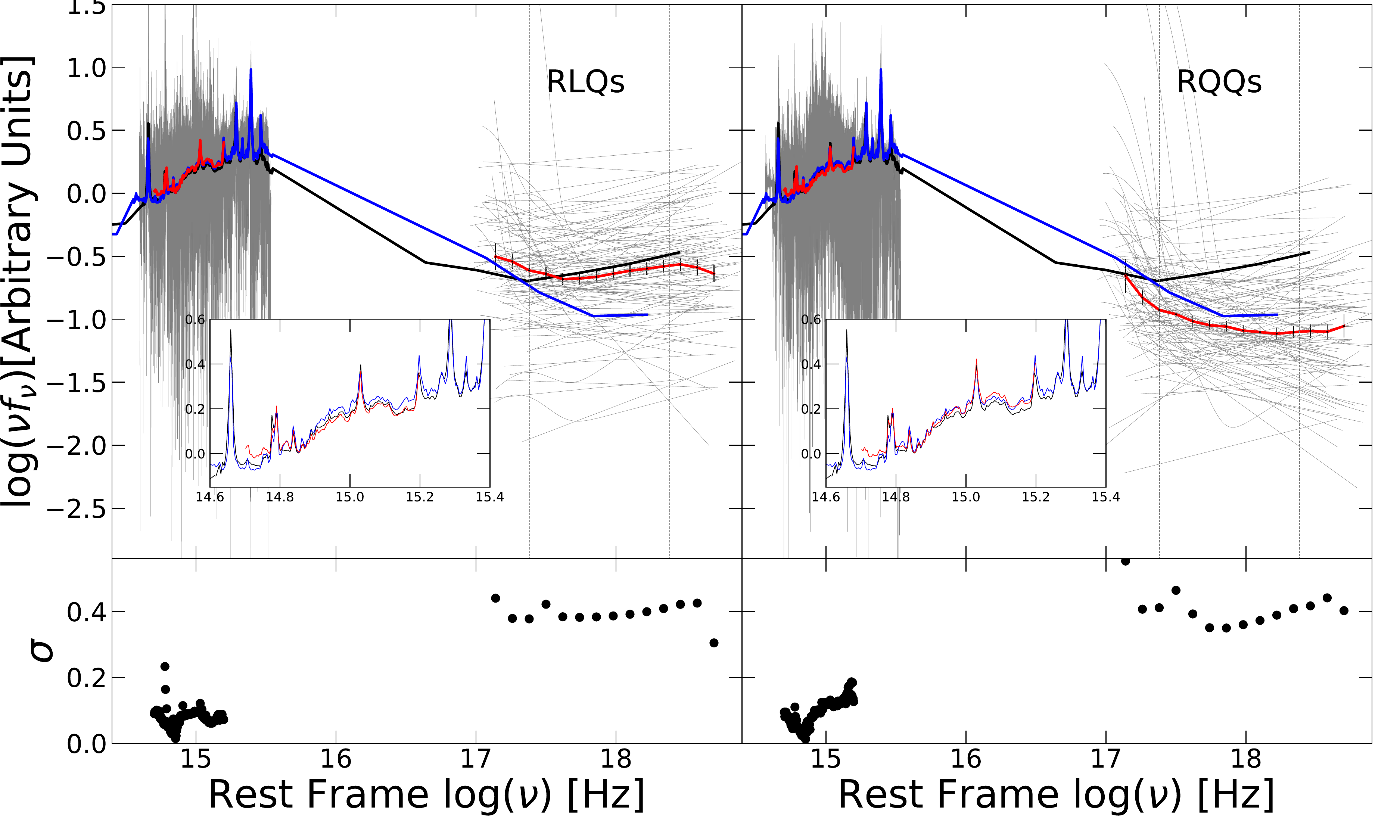}
	\caption{Upper: The median composite SEDs for matched RLQs ($left$) and RQQs ($right$), indicated with red solid lines. Bottom: standard deviation ($\sigma$) around the mean in each wavelength bin. Error bar in X-ray band show the statistical uncertainties ($1.25 \frac{\sigma}{\sqrt{N}}$) of the median values, where $N$ is the number of objects used to construct the SEDs in each wavelength bin. All SEDs of individual quasars are plotted with gray solid lines. The thick black and blue solid lines are composite SEDs for RLQs and RQQs in \cite{2011ApJS..196....2S}, respectively. And the gray dashed lines show the positions of $1\,\rm keV$ and $10\,\rm keV$. The inset shows the details of optical composite spectra. \label{fig:SED}}
\end{figure*}

\begin{figure*}[htb!]
	\centering
	\includegraphics[width=1.0\textwidth]{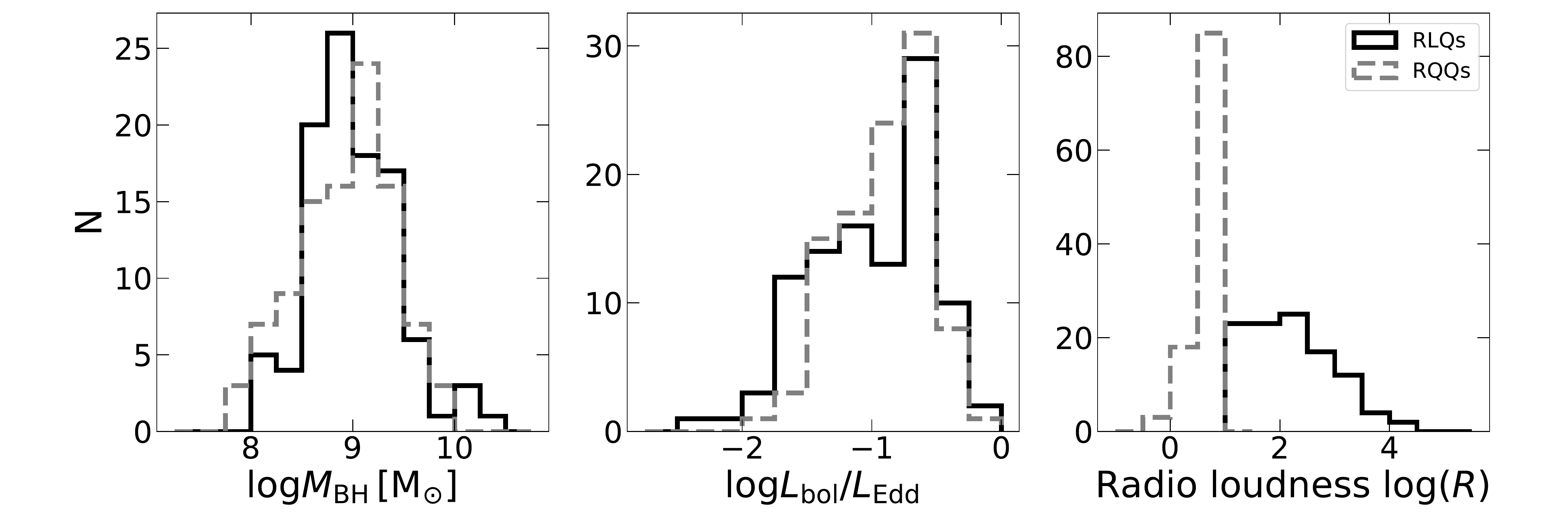}
	\caption{The distributions of black hole mass ($left$), Eddington ratio ($middle$), and radio loudness ($right$) for our matched RLQs and RQQs samples.  \label{fig:prop}}
\end{figure*}

The optical and X-ray composite spectra of RLQs and RQQs are shown in Fig. \ref{fig:SED}. The optical composite spectra of both RLQs and RQQs are very similar to those of S11, and the composite X-ray spectra of both RLQs and RQQs are slightly lower than that of S11. Within the uncertainties of median composite spectra, the composite X-ray spectrum of RLQs is significantly higher than that of RQQs both in the soft and hard X-ray bands, with 0.32 dex and 0.52 dex higher at $1~\rm keV$ and $10~\rm keV$, respectively. The student's $t$-test shows significant differences in the distributions of the normalized X-ray flux at $1~\rm keV$ and $10~\rm keV$ for RLQs and RQQs samples at $>99.9\%$ confidence level, as shown in Fig. \ref{fig:SED} with dashed lines. 
The RLQs and RQQs samples were selected with the same X-ray completeness and matched in $z-\log \nu L_{\nu,\rm2500\AA}$ space as shown with the similar distributions of redshift and the continuum luminosity at rest-frame $2500~\rm \AA$ with $P-$values $>$ 0.999 from KS-test (see Fig. \ref{fig:rl-rq}). 
To further study the difference between RLQs and RQQs samples, we also compared the black hole mass and Eddington ratio for the two populations. The black hole mass, Eddington ratio, and radio loudness of two samples are shown in Fig. \ref{fig:prop}, where the radio loudness of RQQs in sample B are only upper limits. 
We found that there is no significant difference between the distributions of the black hole masses and Eddington ratio for two samples with the $P-$values of 0.48 and 0.16 from KS-test, respectively. The same X-ray completeness, and the similar distributions of the redshift, optical luminosity, black hole mass, and Eddington ratio strongly indicate that the difference of composite X-ray spectra between RLQs and RQQs is intrinsic, not caused by any selection biases. Our finding that the RLQs sample has higher composite X-ray spectrum than RQQs, is in good agreement with previous works \citep[e.g.,][]{1987ApJ...313..596W, 1987ApJ...323..243W, 2011ApJ...726...20M, 2018MNRAS.480.2861G} that RLQs have statistically higher X-ray emission than RQQs.

\begin{figure}[htb!]
	\centering
	\includegraphics[width=1.0\columnwidth]{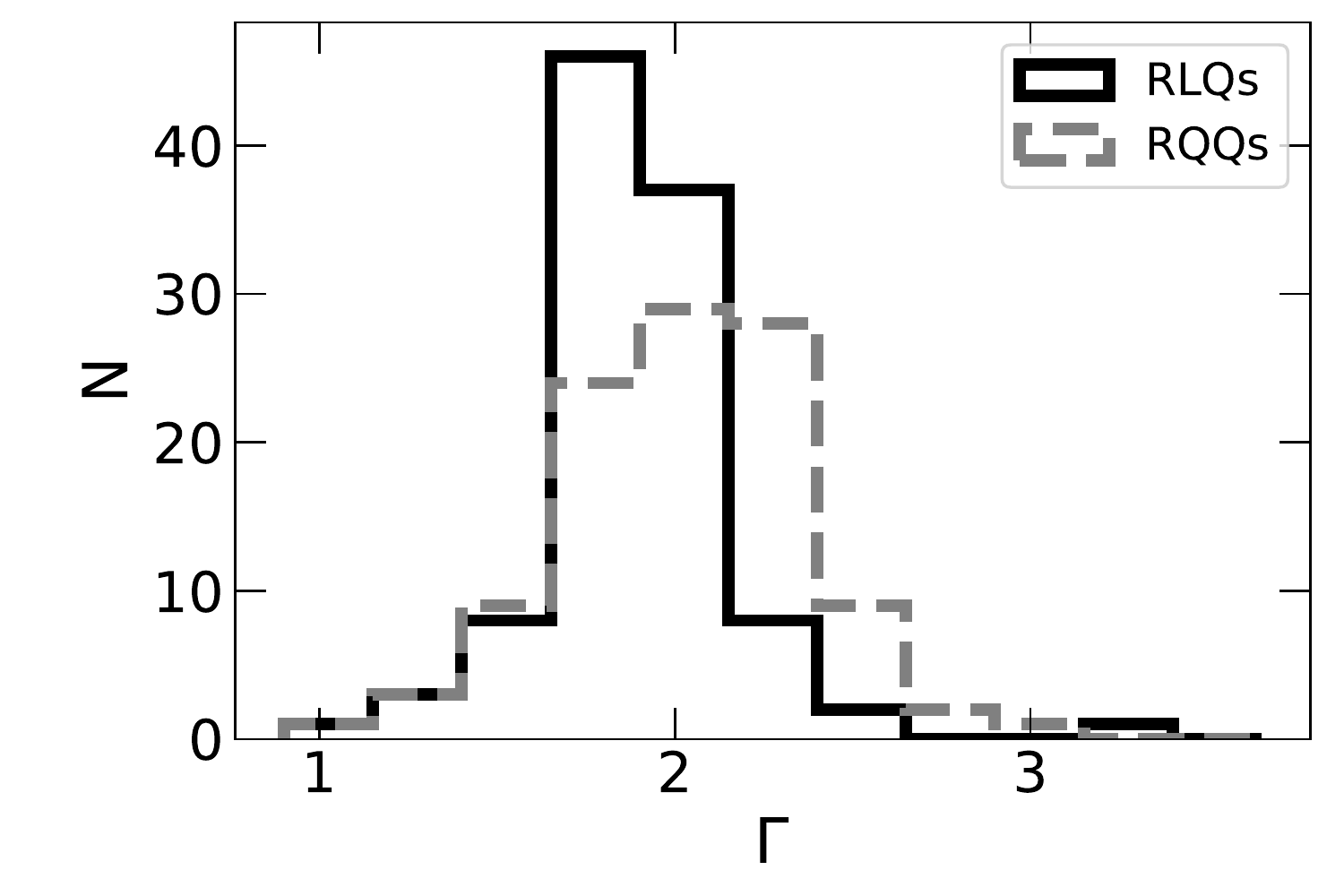}
	\caption{The distributions of X-ray photon index for matched RLQs and RQQs samples. \label{fig:gamma}}
\end{figure}

Fig. \ref{fig:gamma} shows the distributions of the X-ray photon index at $ 2-10\rm \, keV $ for matched RLQs and RQQs samples. The median values of X-ray photon index for RLQs and RQQs are 1.87 and 2.04, respectively. With KS-test, these two distributions show prominent difference with $D = 0.302$ and $P = 8.88\times10^{-5}$. This result is in good agreement with previous works that RLQs have flatter X-ray spectra than RQQs \citep[e.g.,][]{1997MNRAS.288..920L, 1999ApJ...526...60S}.

\section{Discussions} 
\label{sec:discc}

\subsection{Selection biases}

In this work, the X-ray composite spectrum is constructed based on the normalization at 4215 \AA, thus will be affected by any extinction at optical band if present. As shown in \cite{2020ApJsubmitted}, the optical composite spectrum of 3CRR quasars, the radio-selected sample, appears redder compared to S11 quasars, implying the possible extinction at optical/UV band. 
Different from 3CRR quasars, our samples were firstly selected from SDSS, thus are optical-selected samples. 
Therefore, the significant optical extinction in 3CRR quasars will not be expected. 
While most SDSS quasars were color-selected for spectroscopical observations, radio sources have been selected due to their radio detections \cite[e.g.,][]{2002AJ....123.2945R, 2016ApJ...818L...1S}. 
RLQs may have been affected by dust extinction, or have intrinsically redder optical spectra than RQQs. 
Such bias should be carefully addressed to build samples with matched colors \citep[e.g.,][]{2019SCPMA..6269511C}. The composite optical spectrum of RLQs in S11 is slightly redder than that of RQQs. When matching the samples with redshift and optical luminosity, only subtle difference of the optical composite spectra is found between our RLQs and RQQs (see Fig. \ref{fig:SED}). Same with \cite{2019SCPMA..6269511C}, the $g-r$ color of RLQs and RQQs samples show similar distributions with KS-test of $P = 0.823$, indicating that the color selection bias could be negligible for these two samples. Therefore, the composite spectra will not be significantly affected by the normalized luminosity at $4215~\rm \AA$ (or $2500~\rm \AA$), of which the significant extinction is less likely. 

On the other hand, we investigated the presence of significant X-ray absorption in our RLQs and RQQs samples based on the X-ray spectral fitting. There is no heavily absorbed quasar with $N_{\rm H} \ge 10^{23}~\rm cm^{-2}$ in both samples, and only 4 RLQs and 8 RQQs have $N_{\rm H} \ge 10^{22}~\rm cm^{-2}$. Moreover, there is no difference in the X-ray {H\,{\footnotesize I}} column density between RLQs and RQQs samples.
Therefore, the difference in the X-ray composite spectrum between RLQs and RQQs is not affected by the HI absorption. 

It can be clearly seen from Fig. \ref{fig:SED} that the X-ray composite spectrum of RLQs is higher than that of RQQs at hard X-ray band. It is well known that the optical-to-X-ray spectral index depends on the optical/UV luminosity \cite[e.g.,][]{2007ApJ...665.1004J}. Therefore, any systematic difference of optical/UV luminosity distributions between two populations will result in different X-ray spectrum. However, as presented in Fig. \ref{fig:rl-rq}, the similar distributions of optical luminosity between RLQs and RQQs show that this effect is less likely. 

\subsubsection{Redshift effect}
\label{subsec:z}

From comparison studies, \cite{2004AJ....128..523B} found that moderate RLQs ($1<\log R< 2.5$) at $z>4$ have similar X-ray properties to their low-redshift counterparts. In contrast, \cite{2013ApJ...763..109W} and \cite{2019MNRAS.482.2016Z} found that the highly radio-loud quasars (HRLQs, $\log R >2.5$) show an apparent enhancement in the X-ray band at high redshift ($z>4$) and the results can be explained by a fractional inverse-Compton/cosmic microwave background (IC/CMB) model. Moreover, \cite{2013ApJ...763..109W} found that HRLQs at $z>3$ have an X-ray emission enhancement over HRLQs at $z<3$. 

\begin{figure}[htb!]
	\centering
	\includegraphics[width=1.0\columnwidth]{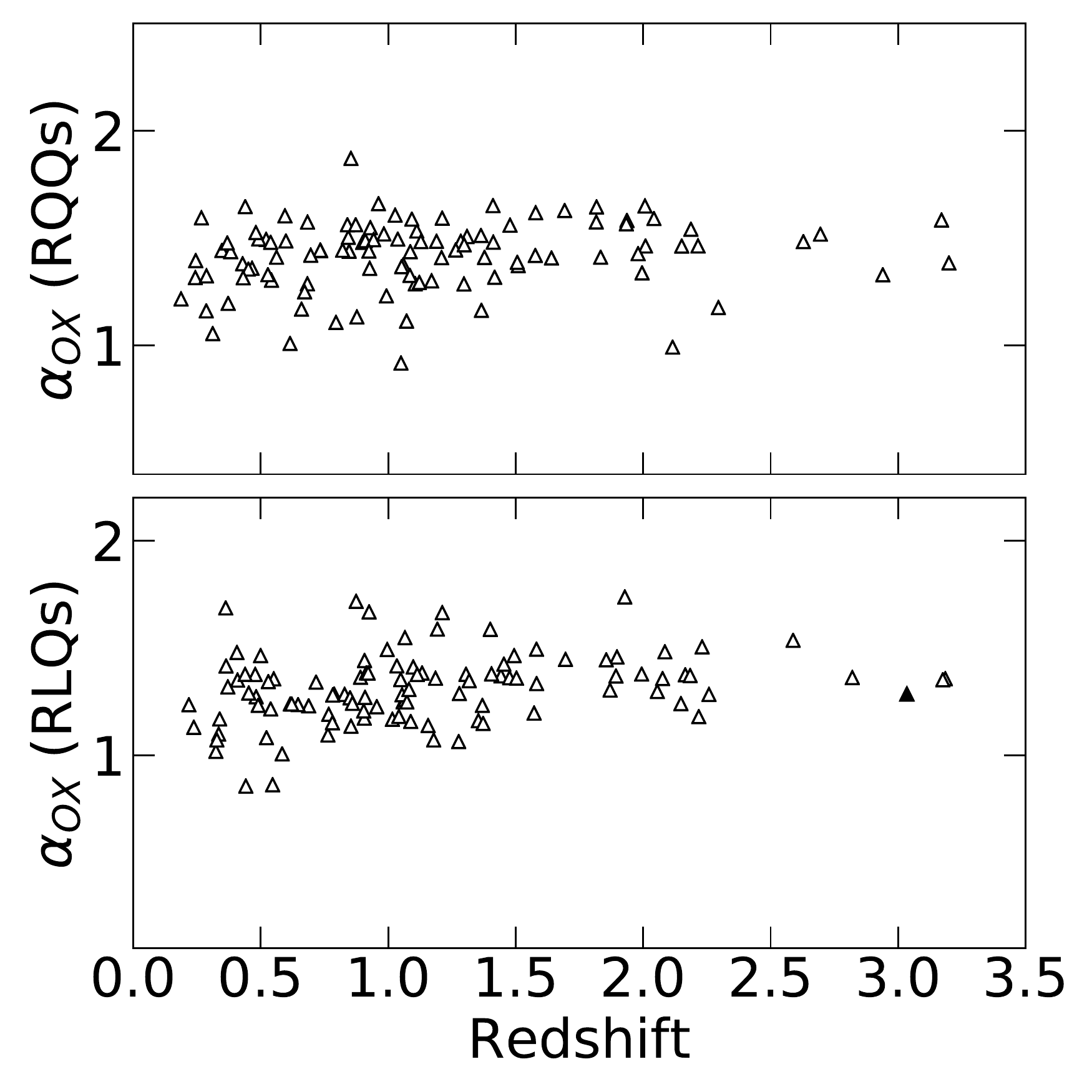}
	\caption{The relation between $\alpha_{\rm ox}$ and redshift for our matched RQQs ($upper$) and RLQs ($lower$) samples. The filled triangle is HRLQ ($ \log R > 2.5 $) at $z>3$. \label{fig:a_z}}
\end{figure}

As Fig. \ref{fig:rl-rq} shows, while there are no RLQs at $z>4$, some RLQs at $ z > 3 $ are indeed included in our sample. To test the effect that high-redshift RLQs may have extra X-ray component (e.g., from IC/CMB), we studied the relationship between the optical-to-X-ray spectral index $\alpha_{\rm ox}$ (defined as $\alpha_{\rm ox} = -\frac{\log(L_{\nu,2\rm keV}/L_{\nu,2500\rm \AA})}{\log(\nu_{2\rm keV}/\nu_{2500})}=-0.384\log(L_{\nu,2\rm keV}/L_{\nu,2500\rm \AA})$) and redshift (see Fig. \ref{fig:a_z}). There are no significant correlations between $ \alpha_{\rm ox} $ and redshift for both RQQs and RLQs. Our result of no significant evolution of $\alpha_{\rm ox}$ with redshift is in good agreement with previous works \citep[e.g.,][]{2007ApJ...665.1004J,2011ApJ...726...20M}. 
RLQs and RQQs have same redshift distribution (see Fig. \ref{fig:rl-rq}). This implies that the difference of the composite X-ray spectra between RLQs and RQQs may not be affected by the redshift.
Furthermore, the $\alpha_{\rm ox}$ of HRLQ at $z>3$ (SDSS J083910.89$ + $200207.3, see Fig. \ref{fig:a_z}) is not different from other RLQs, again supporting that the X-ray enhancement (e.g., due to IC/CMB) is likely not evident. 

\subsection{Radio loudness} \label{subsec:r_loudness}

To further study the difference of the X-ray composite spectra between RLQs and RQQs, the dependence of the X-ray emission on radio loudness is investigated. We divided the quasars in sample A with measured radio loudness into four subsamples with radio loudness $R<10$, $10\le R<100$, $100\le R<1000$, and $R\ge 1000$, which are 
referred as A-0, A-1, A-2, and A-3, respectively. 
The ideal subsamples should be X-ray complete, and have matched redshift, optical luminosity, black hole mass, and also Eddington ratios. While to build such ideal samples is hard in this work, we only assembled subsamples to have same X-ray completeness (adopting the lowest completeness, $\sim 21\%$ in Table \ref{tab:X-comp}) by adjusting the X-ray photon count thresholds. 
In each subsamples, the X-ray photon count thresholds and the number of sources (N) are listed in Table \ref{tab:bins}.

\begin{table*}
	\begin{center}
		\caption[]{The bin used in building X-ray composite spectrum}\label{tab:bins}
		\begin{threeparttable}
		\begin{tabular}{l|crrrrr}
			\hline\noalign{\smallskip}
			\multicolumn{1}{c}{Samples} & \multicolumn{1}{c}{X Photons} &
			\multicolumn{1}{c}{$\rm  log~ \nu_1$} & $\rm  log~ \nu_2$ & bins & $ \Delta \rm  log~ \nu $ & N \\
			\multicolumn{1}{c}{(1)} & \multicolumn{1}{c}{(2)} & \multicolumn{1}{c}{(3)} & \multicolumn{1}{c}{(4)} & \multicolumn{1}{c}{(5)} & \multicolumn{1}{c}{(6)} & \multicolumn{1}{c}{(7)} \\
			\hline\noalign{\smallskip}
			RLQs & 315 & 17.08 & 18.76 & 14 & 0.12 & 106 \\
			RQQs & 200 & 17.08 & 18.76 & 14 & 0.12 & 106 \\
			\hline
			Radio loudness & \multicolumn{6}{c}{Sample A} \\
			\hline
			$R<10$ & 1500 & 17.08 & 18.40 & 11 & 0.12 & 22 \\
			$10\le R<100$ & 275 & 17.20 & 18.64 & 12 & 0.12 & 69 \\
			$100\le R<1000$ & 200 & 17.20 & 18.64 & 12 & 0.12 & 54 \\
			$ R\ge1000$ & 570 & 17.20 & 18.64 & 12 & 0.12 & 17 \\
			\noalign{\smallskip}\hline
		\end{tabular}
	\begin{tablenotes}
		\footnotesize
		\item In this Table, Column (1): Samples; Column (2): The X-ray photon count thresholds; Column (3$-$4): Frequency ranges in $ \log(\rm Hz) $; Column (5): Number of bins in the range; Column (6): Bin size in $ \log(\rm Hz) $; Column (7): Number of quasars in each sample.
	\end{tablenotes}
	\end{threeparttable}
	\end{center}
\end{table*}

\begin{figure}[htb!]
	\centering
	\includegraphics[width=1.0\columnwidth]{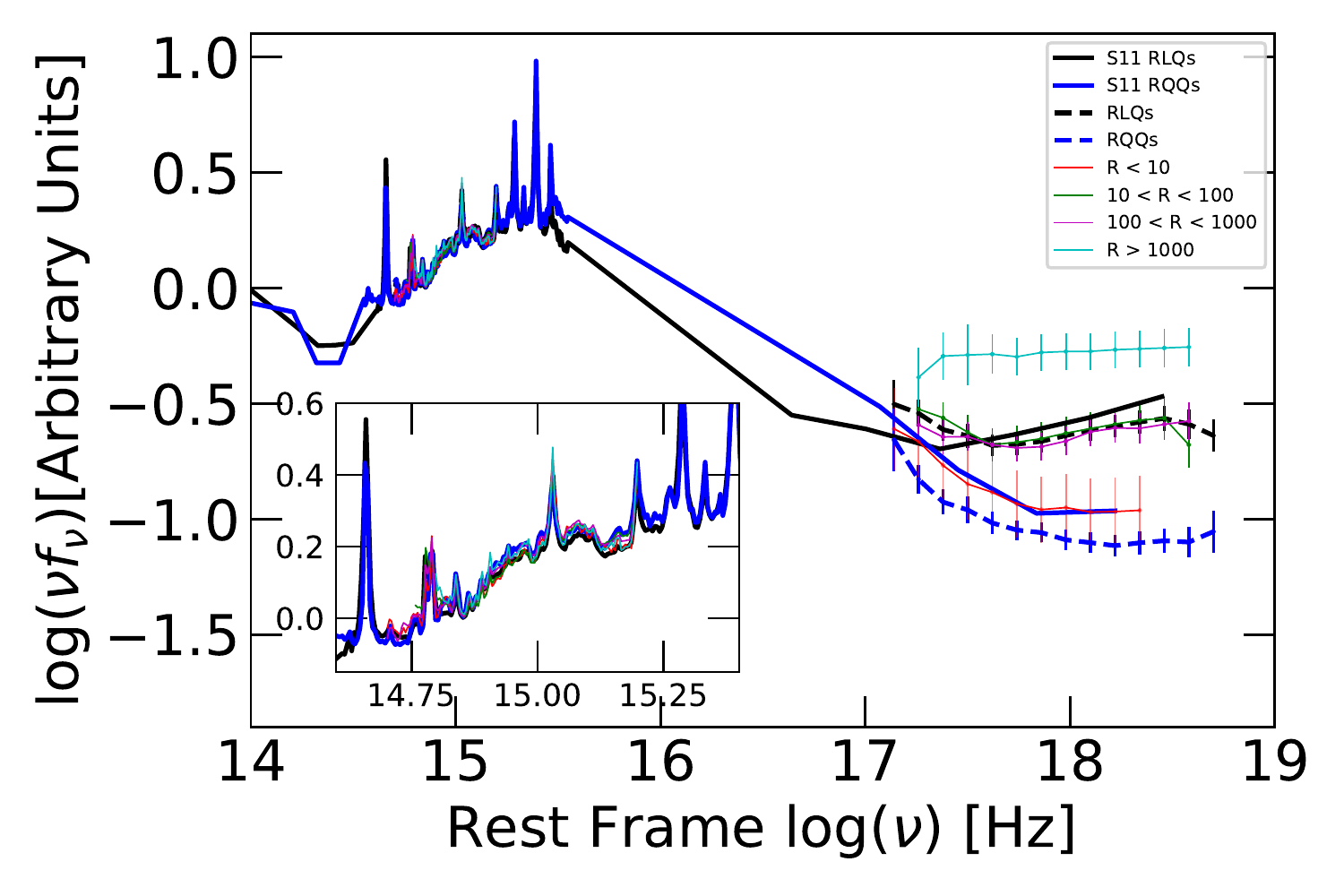}
	\caption{The median composite SEDs of our sample and S11 quasars. The lines are indicated in the upper-right corner.
		The inset shows the details of optical composite spectra.  \label{fig:SEDs}}
\end{figure}

\begin{figure}[htb!]
	\centering
	\includegraphics[width=1.0\columnwidth]{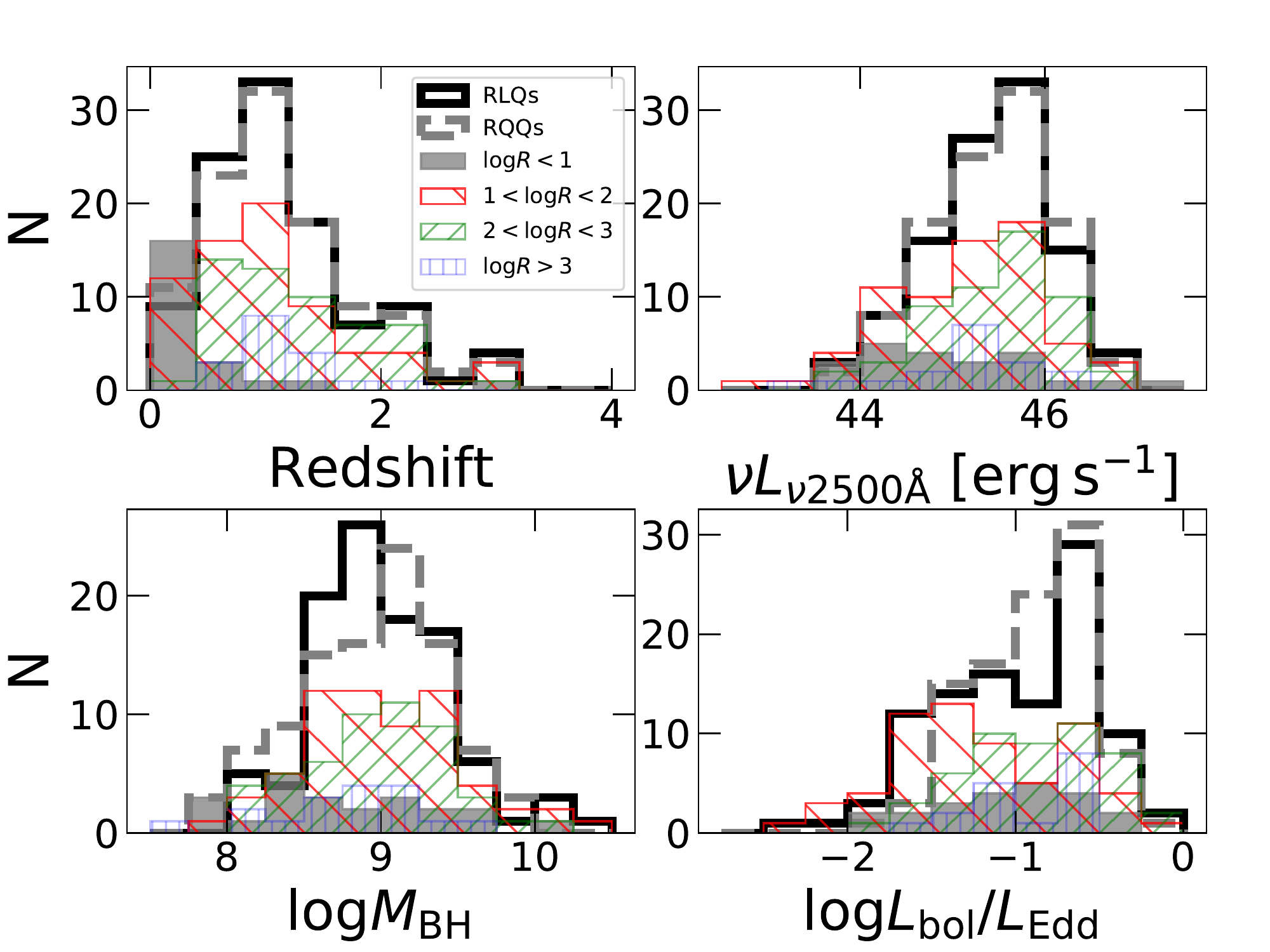}
	\caption{The distributions of redshift ($up-left$), the luminosity at $2500\,\rm \AA$ ($up-right$), black hole mass ($bottom-left$), and Eddington ratio ($bottom-right$) for our subsamples.  \label{fig:sub_prop}}
\end{figure}

\begin{figure}[htb!]
	\centering
	\includegraphics[width=1.0\columnwidth]{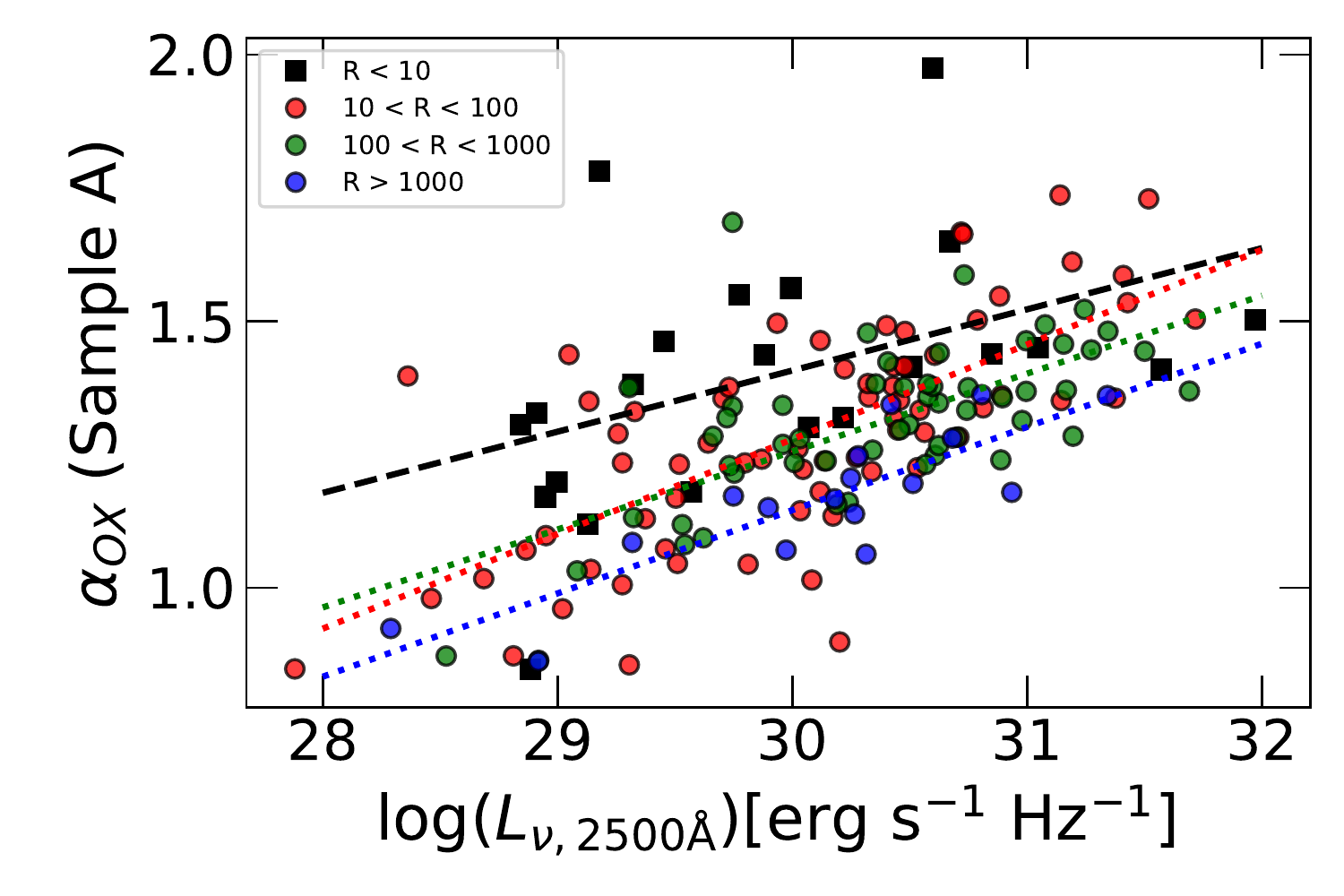}
	\caption{The relationship between $\alpha_{\rm ox}$ and $L_{\nu,2500\rm \AA}$ in radio loudness bins. The best linear fit and sources in each bin are shown with same color.  \label{fig:aox_r}}
\end{figure}

We constructed the optical and X-ray composite spectra for these subsamples with the same method as in Section \ref{sec:comp} with different frequency range and bin number at X-ray bands as shown in Table \ref{tab:bins}. The X-ray composite spectra are shown in Fig. \ref{fig:SEDs}. While the optical composite spectra of these subsamples are almost same, the X-ray spectra are quite different, which however needs to be further investigated considering the distributions of the redshift, optical luminosity, black hole mass, and Eddington ratio (see Fig. \ref{fig:sub_prop}). We found that the A-2 and A-3 show similar redshift, optical luminosity, black hole mass, and Eddington ratio distributions with all KS-test $P-$values greater than 0.30. 
Not influenced by these effects, the A-3 has a higher X-ray composite spectrum than A-2, as shown in Fig. \ref{fig:SEDs}. 
It may indicate that X-ray emission is higher when radio loudness is higher. While this trend is not seen when compare A-1 and A-2, we found that the redshift and optical luminosity of A-1 are systematically lower than A-2. Most likely due to the relation between $\alpha_{\rm ox}$ and optical/UV luminosity \citep[e.g.,][see also Fig. \ref{fig:aox_r}]{2011ApJ...726...20M}, the higher X-ray spectrum in A-1 may be caused by their systematic low optical luminosity. We tried to re-construct the composite X-ray spectra for A-1 and A-2 by matching the redshift and optical luminosity. Indeed, the composite X-ray spectrum of A-1 is lower than that of A-2, however only slightly. The small source number precludes us to draw firm conclusions. 
The X-ray composite spectrum of A-0 is lower than that of A-2. We noticed that sample A-0 has lower redshift (mostly $z<0.4$) and optical luminosity than A-2 (see Fig. \ref{fig:sub_prop}). Considering the optical/UV luminosity $-$ $\alpha_{\rm ox}$ relation, the X-ray composite spectrum of sample A-0 will be even lower than it appears in Fig. \ref{fig:SEDs}. Therefore, our analysis shows the X-ray spectrum tends to be stronger with increasing radio loudness, at least for subsamples A-0, A-2, and A-3.

\begin{table}
	\begin{center}
		\caption[]{The relation of $\alpha_{\rm ox}$ and $L_{\nu,2500\rm \AA}$}\label{tab:linefitbins}
		\begin{tabular}{l|rrrrc}
			\hline\noalign{\smallskip}
			\multicolumn{1}{c}{Sample} & \multicolumn{5}{c}{$\alpha_{\rm ox}=a+b*\log L_{\nu,2500\rm \AA}$}  \\
			\multicolumn{1}{c}{ } & 
			\multicolumn{1}{c}{$ a $} & \multicolumn{1}{c}{$\Delta a$} & \multicolumn{1}{c}{$ b $} & \multicolumn{1}{c}{$ \Delta b $} & \multicolumn{1}{c}{$r_{\rm s}$} \\
			\hline\noalign{\smallskip}
			RLQs & -3.260 & 0.570 & 0.151 & 0.018 & 0.62 \\
			RQQs & -2.375 & 0.630 & 0.125 & 0.021 & 0.53  \\
			\hline
			Radio loudness & \multicolumn{5}{c}{Sample A} \\
			\hline
			$R<10$ & -2.036 & 1.547 & 0.115 & 0.052 & 0.55 \\
			$10\le R<100$ & -4.052 & 0.677 & 0.178 & 0.022 & 0.69 \\
			$100\le R<1000$ & -3.132 & 0.706 & 0.146 & 0.023 & 0.62 \\
			$ R\ge1000$ & -3.540 & 0.784 & 0.156 & 0.026 & 0.76 \\
			\noalign{\smallskip}\hline
		\end{tabular}
	\end{center}
\end{table}

The tendency of higher X-ray spectra with increasing radio loudness can be alternatively explored when we plot the relationship between $\alpha_{\rm ox}$ and $L_{\nu,2500\rm \AA}$ in the subsamples (see Fig. \ref{fig:aox_r}). 
It can be seen that the optical-to-X-ray slope $\alpha_{\rm ox}$ tends to be flatter when the radio loudness increases, albeit large scatters. At fixed optical luminosity $L_{\nu,2500\rm \AA}$, $\alpha_{\rm ox}$ is anti-correlated with radio loudness. 
This is strongly supported by the comparison between samples A-2 and A-3, with similar redshift and optical luminosity distributions. The subsample A-3 has statistically lower $\alpha_{\rm ox}$ than that of A-2, with student's $t$-test at $>99\%$ confidence. 
In each subsample, we used the ordinary least squares (OLS) linear relation $\alpha_{\rm ox}=a+b*\log L_{\nu,2500\rm \AA}$ to fit the data. The fitting results are shown in Table \ref{tab:linefitbins}, in which the Spearman correlation coefficient $r_{\rm s}$ are also given for the significant correlations all at $\ge 99.9\%$ confidence level.

\subsubsection{X-ray excess} 
\label{subsubsec:aox}

\begin{figure}[htb!]
	\centering
	\includegraphics[width=1.0\columnwidth]{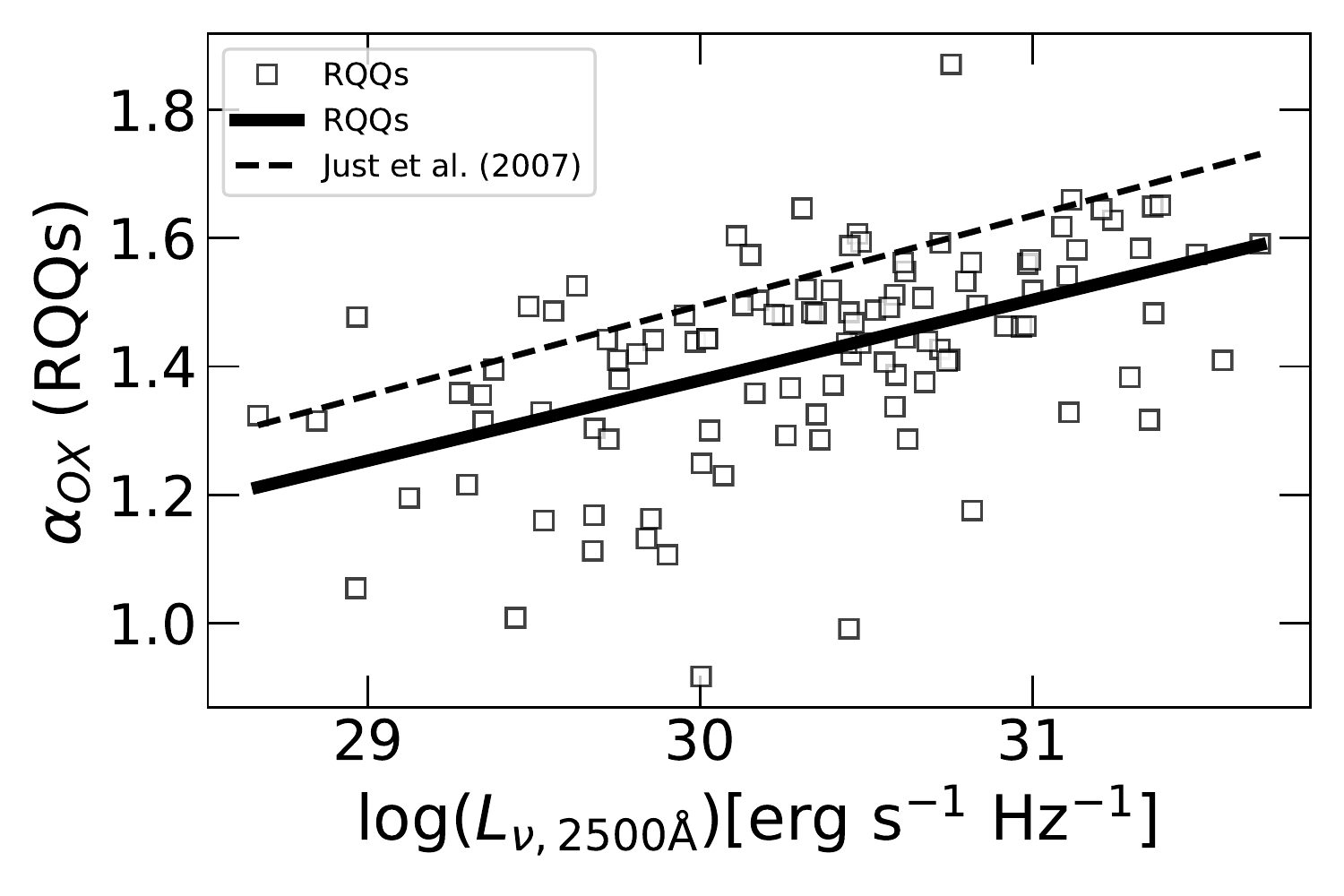}
	\caption{The relationship between $\alpha_{\rm ox}$ and $L_{\nu,2500\rm \AA}$ for RQQs. The black solid line shows the best linear fit. The dashed line is the relation reported in \cite{2007ApJ...665.1004J}. \label{fig:aox_rqq}}
\end{figure}

As did in \cite{2011ApJ...726...20M} and \cite{2013ApJ...763..109W}, we investigate the X-ray excess of radio-detected quasars in sample A (i.e., RD-RQQs and RLQs) relative to RQQs, which is defined as
\begin{equation}
\Delta\alpha_{\rm ox}=\alpha_{\rm ox}-\alpha_{\rm ox,RQQ}
\end{equation}
with $\alpha_{\rm ox,RQQ}$ being the expected value calculated from $\alpha_{\rm ox}-L_{\nu,2500\rm \AA}$ relation of RQQs.
In Fig. \ref{fig:aox_rqq}, we plot the relation of $\alpha_{\rm ox}$ with $L_{\nu,2500\rm \AA}$ for our RQQs sample. The Spearman correlation analysis shows a significant correlation with coefficient of $r_{\rm s}=0.53$ at $> 99.99\%$ confidence level. 
The best OLS linear fit gives
\begin{equation}
\alpha_{\rm ox}=(0.125\pm0.021)\log(L_{\nu,2500\rm \AA})-(2.375\pm0.630),
\label{eq:aox_L2500}
\end{equation}
which is listed in Table \ref{tab:linefitbins}. 
This relation is consistent with $\alpha_{\rm ox}=(0.140\pm0.007)\log(L_{\nu,2500\rm \AA})-(2.705\pm0.212)$ for a sample of RQQs reported in \cite{2007ApJ...665.1004J}, which is also plotted in Fig. \ref{fig:aox_rqq}. 
With a note, the X-ray completeness of our RQQs sample is about 0.19 which is much lower than that of \cite{2007ApJ...665.1004J} sample (about 0.89). The lower X-ray completeness in our RQQs sample may selected brighter X-ray sources than \cite{2007ApJ...665.1004J} and caused the statistically lower values of $ \alpha_{\rm ox} $ for our RQQs in Fig. \ref{fig:aox_rqq}.
In this work, same as \cite{2013ApJ...763..109W}, we used the $\alpha_{\rm ox} - \log (L_{\nu,2500\rm \AA})$ relation of \cite{2007ApJ...665.1004J} to estimate $\alpha_{\rm ox,RQQ}$, and then $\Delta\alpha_{\rm ox}$ was calculated. 

\begin{figure}[htb!]
	\centering
	\includegraphics[width=1.0\columnwidth]{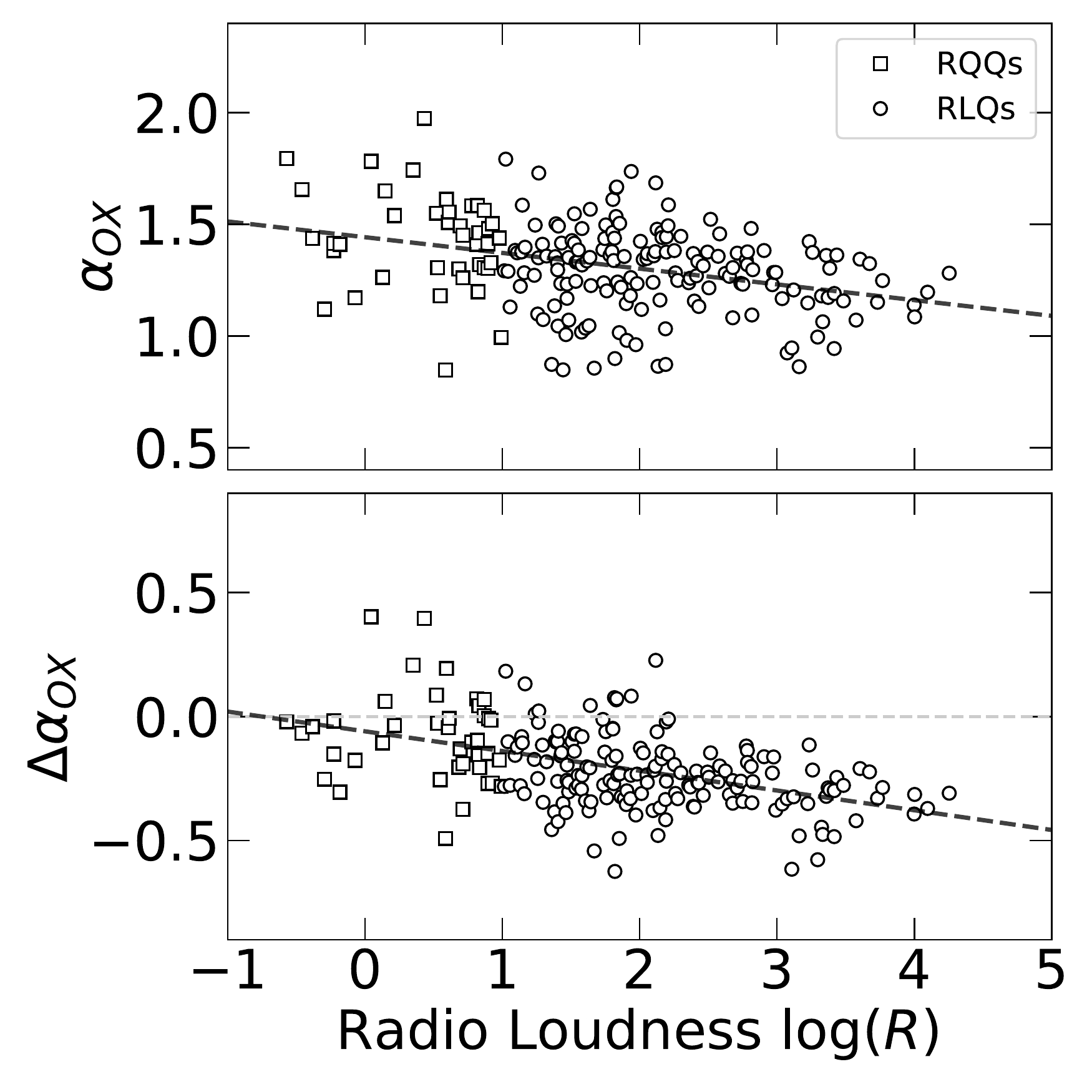}
	\caption{The relationship between radio loudness and $\alpha_{\rm ox}$ or $\Delta\alpha_{\rm ox}$ in sample A quasars. The black dashed lines show the best-fit relation. 
	The gray dashed line presents the position of $\Delta\alpha_{\rm ox}=0$. \label{fig:aox_r_bin}}
\end{figure}


In Fig. \ref{fig:aox_r_bin}, the relationships between the radio loudness and $\alpha_{\rm ox}$ or $\Delta\alpha_{\rm ox}$ are presented for radio-detected quasars with X-ray photons greater than 200 in Sample A. There is a strong negative correlation between $\alpha_{\rm ox}$ and radio loudness with a Spearman correlation coefficient $r_{\rm s}=-0.33$ at $>99.99\%$ confidence level, indicating smaller $\alpha_{\rm ox}$ (i.e., flatter optical-to-X-ray spectral slope) with increasing radio loudness. Interestingly, RD-RQQs follow the general trend. 
The linear fit between $\alpha_{\rm ox}$ and $\log R$ gives 
\begin{equation}
\alpha_{\rm ox}=(-0.070\pm0.014)\log R+(1.442\pm0.029).
\end{equation}

Based on the definition, $\Delta\alpha_{\rm ox}$ quantifies the excess of X-ray emission in RLQs compared to that of RQQs, in which the X-ray emission is thought to be mainly from disk-corona system \citep{2013ApJ...763..109W}. The excess of X-ray emission in RLQs may come from relativistic jets as argued by \cite{2011ApJ...726...20M} and \cite{2013ApJ...763..109W}. 
We found a significant positive correlation between the excess of X-ray emission and radio loudness in RLQs with a Spearman correlation coefficient $r_{\rm s}=0.47$ at $>99.99\%$ confidence level, as shown in Fig. \ref{fig:aox_r_bin}. 
We linearly fitted the relation between $\Delta\alpha_{\rm ox}$ and $\log R$, and found 
\begin{equation} 
\label{eqa:deltaox_r}
\Delta\alpha_{\rm ox}=(-0.080\pm0.010)\log R-(0.060\pm0.022).
\end{equation}

The negative values of $\Delta\alpha_{\rm ox}$, and its strong correlation with radio loudness in RLQs may support the presence of jet-linked X-ray emission, which is likely higher at radio-louder sources due to normally more powerful jets.

\subsubsection{Soft X-ray excess} \label{subsec:softX-ray}

As shown in Fig. \ref{fig:SED}, our RQQs sample has lower X-ray composite spectrum than that of S11 RQQs. In contrast, the composite X-ray spectrum of our RLQs has slightly lower hard X-ray flux than S11 RLQs, however higher at soft X-ray band. Moreover, the soft X-ray composite spectrum of our RLQs is higher than that of RQQs, which however is absent between S11 RLQs and RQQs.

The X-ray spectral analysis shows that out of 103 NRD-RQQs, 44 objects show prominent soft X-ray excess. The detection rate of soft X-ray excess in these NRD-RQQs is about $42.7\%$. In radio-detected quasars of sample A, the detection rate is about $60.0\%$ and $25.0\%$ for 40 RD-RQQs and 160 RLQs, respectively. The results are in agreement with previous works that RLQs have less fraction of detected soft X-ray excess than RQQs \citep[e.g.,][]{Scott2011MNRAS, Boissay2016AandA}. 

\subsection{Radio$-$X-ray relation} 
\label{subsec:dichotomy}

\begin{figure}[htb!]
	\centering
	\includegraphics[width=1.0\columnwidth]{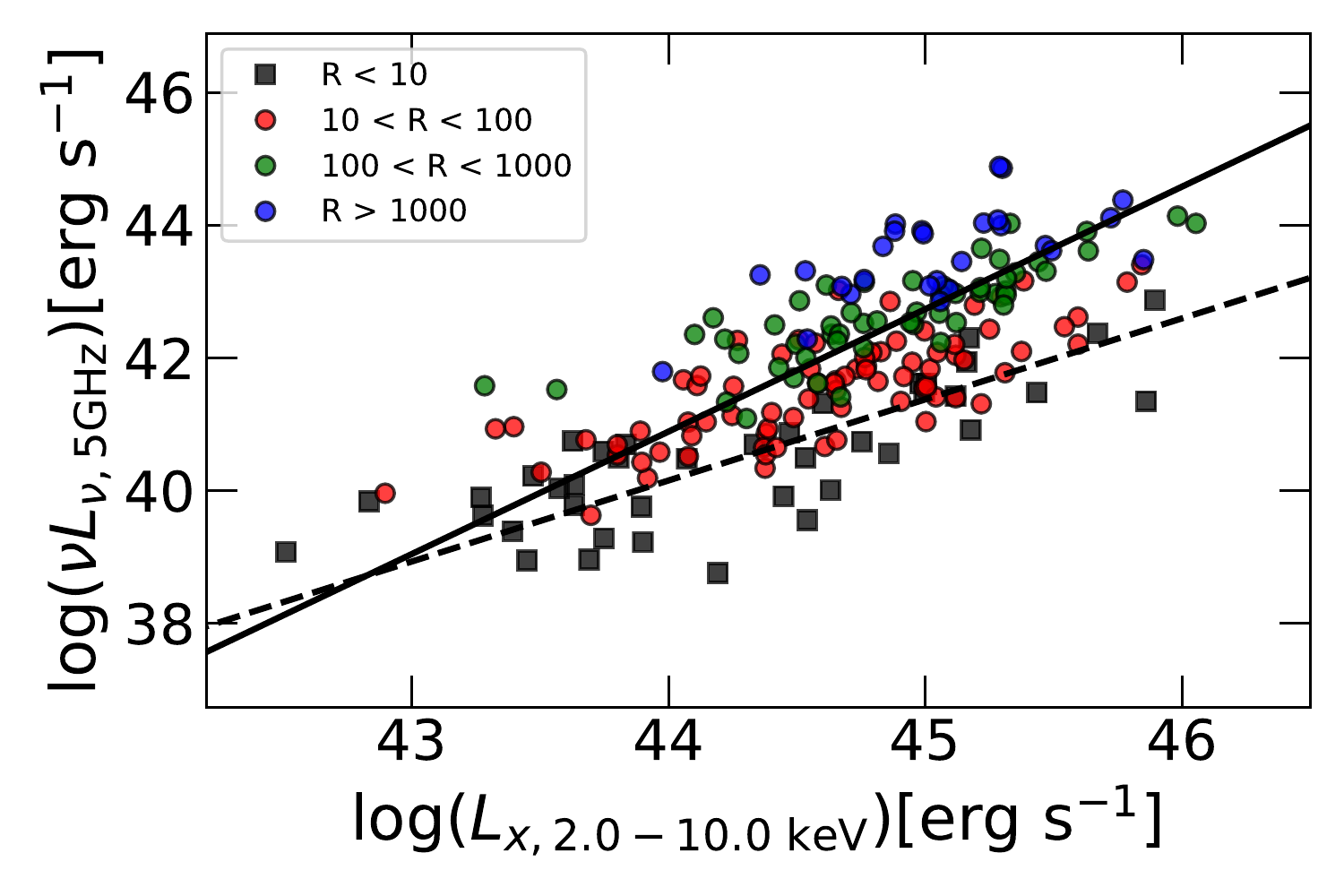}
	\caption{The relationship between the radio and X-ray luminosities. The solid and dashed lines are the linear fits for RLQs and RD-RQQs, respectively. \label{fig:r_x}}
\end{figure}

The radio$ - $X-ray correlation has been widely studied in many occasions, which was used to study the mechanism of X-ray emission, either from accretion flow or relativistic jet, while the radio emission is usually thought to be from jet \cite[e.g.,][]{2003MNRAS.345.1057M,2004A&A...414..895F,2006ApJ...645..890W,2008ApJ...688..826L,2018MNRAS.481L..45L}. We present the relationship between the radio and X-ray luminosities for radio-detected quasars in Fig. \ref{fig:r_x}, and find a significant correlation between two parameters. After excluding the common dependence on redshift, the radio and X-ray luminosity still show a strong positive correlation with partial Spearman correlation coefficient of 0.253 at $>99.9\%$ confidence. The OLS bisector linear fit shows $\log L_r = (1.846\pm 0.103) \log L_X - (40.358\pm 4.621)$ for all RLQs, and $\log L_r = (1.222\pm 0.113) \log L_X - (13.626\pm 5.014)$ for RD-RQQs. The steeper index in RLQs than RD-RQQs is consistent with the results of \cite{2006ApJ...645..890W} and \cite{2008ApJ...688..826L}. 

\begin{table}
	\begin{center}
		\caption[]{The multiple linear fit in the sample}\label{tab:r_x_R}
		\begin{tabular}{l|rrr}
			\hline\noalign{\smallskip}
			\multicolumn{1}{c}{Sample} & \multicolumn{3}{c}{$\log L_r=a*\log L_x+b*\log (R)+c$} \\
			\multicolumn{1}{c}{ } & \multicolumn{1}{c}{$ a $} & \multicolumn{1}{c}{$ b $} & \multicolumn{1}{c}{$ c $} \\
			\hline\noalign{\smallskip}
			RLQs & $0.97\pm 0.15$ & $0.86\pm 0.11$ & $-3.25\pm 6.51$ \\
			RD-RQQs & $0.99\pm 0.19$  & $0.75\pm 0.35$ & $-3.73\pm 8.57$ \\
			\hline
			$10\le R<100$ & $0.92\pm 0.20$ & $0.98\pm 0.43$ & $-1.13\pm 8.84$ \\
			$100\le R<1000$ & $1.04\pm 0.26$ & $0.80\pm 0.47$ & $-5.94\pm 11.69$ \\
			$ R\ge1000$ & $1.02\pm 0.45$ & $1.29\pm 0.60$ & $-6.94\pm 20.35$ \\
			\hline
			All & $0.98\pm 0.12$ & $0.81\pm 0.08$ & $-3.46\pm 5.17$ \\
			\noalign{\smallskip}\hline
		\end{tabular}
	\end{center}
\end{table}

\begin{figure}
	\centering
	\includegraphics[width=1.0\columnwidth]{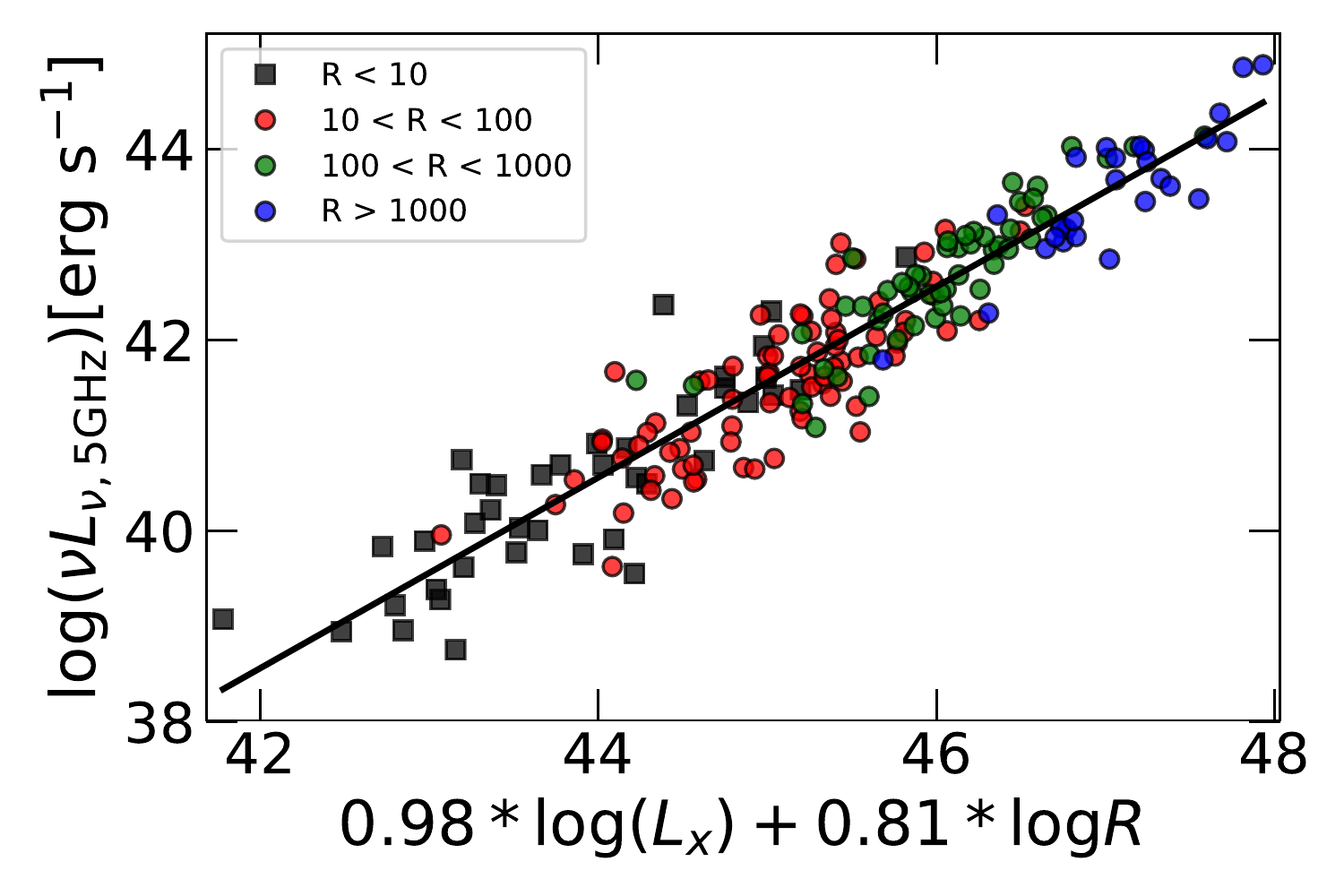}
	\caption{The dependence of the radio luminosity on the X-ray luminosity and radio loudness in all radio-detected quasars. \label{fig:r_x_R}}
\end{figure}

It is clearly seen from Fig. \ref{fig:r_x} that the radio luminosity at fixed X-ray luminosity increases with increasing radio loudness, and the sources in different radio loudness bins may have different radio$-$X-ray relations.
The dependence of the radio luminosity at $5~\rm GHz$ ($L_{\rm r}$) on the X-ray luminosity ($L_{\rm X}$) and radio loudness ($R$) can be investigated from multiple linear fit by Markov chain Monte Carlo (MCMC) method with the python-based $emcee$ code \citep{2013PASP..125..306F}. The results are shown in Table \ref{tab:r_x_R} for all quasars and subsamples in sample A. When adding radio loudness, we found that the radio$-$X-ray index is about 1.0 for all radio-detected quasars, RD-RQQs, RLQs, and RLQs subsamples.
The unified relationship between $L_{\rm r}$, $L_{\rm X}$ and $R$ for all radio-detected quasars is shown in Fig. \ref{fig:r_x_R}.

\subsection{The mechanism of X-ray emission}
\label{subsec:X-emission}

After fully excluding the selection biases caused by the X-ray completeness, the distributions of redshift, optical luminosity, black hole mass, and Eddington ratio, we robustly found that the X-ray composite spectrum of RLQs is higher than that of RQQs when normalized at rest frame 4215\AA. Similar to our work, \cite{2012A&A...545A..66B} explored also the interplay between radio and X-ray emission in AGNs with SDSS, XMM-Newton, and FIRST observations. They claimed that radio-loud AGNs are also X-ray loud, with higher X-ray-to-optical ratio than that of radio-quiet objects. It agrees with our result that RLQs have higher X-ray composite spectrum than that of RQQs. 
With hardness ratio analysis, \cite{2012A&A...545A..66B} found the flatter photon index in radio-loud sources than radio-quiet objects. 
Moreover, they also found a tight relationship between X-ray and radio luminosity, i.e., the  higher X-ray luminosity corresponds to higher radio luminosity. 
All these results found in \cite{2012A&A...545A..66B} are consistent with ours. 

Hard X-ray emission in radio-quiet AGN is believed to be produced via inverse Compton
scattering by hot and compact corona near the supermassive black hole, however, its origin and physical properties, including geometry, kinematics, and dynamics, are still unclear \citep[e.g.,][]{1980A&A....86..121S, 2015A&ARv..23....1B}. In addition to corona, strong relativistic jets could produce extra X-ray emission in radio-loud AGNs \cite[e.g.,][]{1987ApJ...313..596W,2011ApJ...726...20M}.

From the comparison of the X-ray emission between Seyfert 1 and Compton-thin Seyfert 2 galaxies, \cite{2014ApJ...783..106L} found that the corona X-ray emission is intrinsically anisotropic, which can be explained by a bipolar outflowing corona with a bulk velocity of $\sim 0.3-0.5~ c$. Their results seem to favor the scenario that the role of the corona could be subsumed by the base of the jet in AGNs \citep{2005ApJ...635.1203M}. Likely, the launches of corona and relativistic jets are directly related. 
As indicated in \cite{2004ApJ...609..972M}, the base of the jet is located within a few tens of gravitational radii and is accelerated along a region of $100-1000\ R_{\rm g}$, where $R_{\rm g}=GM/c^2$. The authors found the velocity of the base of jet is $\sim 0.3-0.4\ c$, and the X-ray emission is dominated by synchrotron self-Compton at the base of the jet. Recently, \cite{2017ApJ...835..226K} studied the jet and coronal properties of AGN, and found that the radio Eddington luminosity inversely scales with X-ray reflection fraction, and positively scales the path length connecting the corona and reflecting regions of the disk. In the corona-jet model proposed by \cite{2004ApJ...609..972M}, the correlations can be explained via a moving corona that is propagating into the large-scale jets.

The radio-detected quasars in our work show a general correlation between $\alpha_{\rm ox}$ and radio loudness, and a unified multi-correlation between $L_{\rm r}$, $L_{\rm X}$ and $R$. These results can be well explained with the corona-jet model, in which the corona and jet are directly related as proposed in \cite{2005ApJ...635.1203M} and \cite{2017ApJ...835..226K}. Normally, the quasars at higher radio loudness tend to have more powerful jets, therefore more luminous X-ray emission will be expected in corona-jet model. 
Indeed, the jet-like radio morphologies are often detected in radio-quiet AGNs \cite[e.g.,][]{2013MNRAS.432.1138P}, but they are much weaker and more compact.

\section{Summary} \label{sec:sum}

We have studied the X-ray emission in SDSS quasars by constructing the optical and X-ray composite spectra normalized at rest frame $4215\rm\, \AA$ (or $2200\,\rm \AA$). We found that the X-ray composite spectrum of RLQs is higher than that of RQQs at $z<3.2$ with both samples matched in X-ray completeness, redshift, and optical luminosity. The soft X-ray excess is evident in both RLQs and RQQs samples. However, the source fraction of detected soft X-ray excess in RLQs is lower than RQQs. We find the X-ray composite spectra are higher with increasing radio loudness. Moreover, a significant correlation is found between $\alpha_{\rm ox}$ and radio loudness, and there is a unified multi-correlation between the radio, X-ray luminosities and radio loudness in radio-detected quasars. These results could be possibly explained with the corona-jet model.

\begin{acknowledgements}
We thank the anonymous referee for valuable and insightful suggestions that improved the manuscript.
We thank Mai Liao, Muhammad Shahzad Anjum, Jiawen Li, Shuhui Zhang and Shuai Feng for usefull discussions.
This work is supported by the National Science Foundation of China (grants 11873073, U1531245, 11773056 and U1831138).
This work is based on results from the enhanced XMM-Newton spectral-fit database, an ESA PRODEX funded project, based in turn on observations obtained with XMM-Newton, an ESA science mission with instruments and contributions directly funded by ESA Member States and NASA. 
Funding for SDSS-III has been provided by the Alfred P. Sloan Foundation, the Participating Institutions, the National Science Foundation, and the U.S. Department of Energy Office of Science. The SDSS-III web site is http://www.sdss3.org/. 
\end{acknowledgements}

\appendix                  

%

\label{lastpage}

\end{document}